\documentclass[a4paper,11pt]{article}

\usepackage{jcappub}
\usepackage{xcolor}
\usepackage[normalem]{ulem}
\usepackage{booktabs}

\usepackage[compat=1.1.0]{tikz-feynman}
\tikzfeynmanset{warn luatex=false}

\newcommand{\cannibalvertices}[5]{%
  \coordinate (u) at (0,0.58);
  \coordinate (d) at (0,-0.58);

  \coordinate[label=left:{$#1$}]  (i1) at (-1.65,1.15);
  \coordinate[label=left:{$#2$}]  (i2) at (-1.65,0);
  \coordinate[label=left:{$#3$}]  (i3) at (-1.65,-1.15);
  \coordinate[label=right:{$#4$}] (o1) at (1.65,0.58);
  \coordinate[label=right:{$#5$}] (o2) at (1.65,-0.58);
}

\newcommand{\darkscalarline}[2]{%
  \draw[/tikzfeynman/scalar, line width=0.55pt] (#1) -- (#2);
}

\newcommand{\darkphotonline}[2]{%
  \draw[/tikzfeynman/photon, line width=0.55pt] (#1) -- (#2);
}

\title{Inverse phase transitions via\\ dark sector chemical equilibration}

\author[a]{Esau Cervantes,}
\author[b,c]{Felix Kahlhoefer,}
\author[b]{Jonas Matuszak,}
\author[b]{and Santiago Rosellon}

\affiliation[a]{National Centre for Nuclear Research,\\
Pasteura 7, 02-093 Warsaw, Poland}

\affiliation[b]{Institute for Theoretical Particle Physics (TTP),\\
Karlsruhe Institute of Technology, 76128 Karlsruhe, Germany}

\affiliation[c]{Institute for Astroparticle Physics (IAP), Karlsruhe Institute of Technology (KIT),
Hermann-von-Helmholtz-Platz 1, 76344 Eggenstein-Leopoldshafen, Germany}

\emailAdd{esau.cervantes@ncbj.gov.pl}
\emailAdd{felix.kahlhoefer@kit.edu}
\emailAdd{jonas.matuszak@kit.edu}
\emailAdd{sroselloni@gmail.com}

\abstract{
We study the cosmological evolution of the effective potential in a dark sector with a spontaneously broken Abelian gauge symmetry, populated out of equilibrium through the freeze-in mechanism. Initially dilute and in the broken phase, the dark sector is described by a temperature as well as a chemical potential that reflects the low occupancy of dark sector states. We show that the thermal corrections to the effective potential depend on both of these quantities, and that they are suppressed while the sector remains dilute, but grow as internal number-changing reactions increase the occupancy and drive the system toward chemical equilibrium. We solve the coupled evolution of the temperature and chemical potential and the resulting impact on the effective potential to show that, for sufficiently large portal couplings, chemical equilibration can evolve the effective potential across a first-order barrier and drive the dark sector toward its symmetric phase through an inverse phase transition.
}


\begin{document}

\noindent{\hfill Preprint: TTP26-037, P3H-26-078}

\vspace{-1.2em}
\maketitle
\raggedbottom

\section{Introduction}
\label{sec:introduction}

First-order phase transitions have received much attention recently as cosmological sources of gravitational waves that can be detected with current and upcoming observations, for example from the PTA collaborations~\cite{NANOGrav:2023gor,EPTA:2023fyk,Reardon:2023gzh,Xu:2023wog,Miles:2024seg}, LISA~\cite{Caprini_2016} and the Einstein Telescope~\cite{Sathyaprakash:2012jk}, but also as a way to produce the relic abundance of dark matter~\cite{Baker:2019ndr,Azatov:2021ifm}. In this context, the system under consideration typically   starts in a symmetric phase at high temperatures and enters a broken phase as the universe cools~\cite{Athron_2024,Caprini_2016,Hall:2019rld}. However, dark sectors with very feeble couplings to the Standard Model (SM) plasma allow for a wider range of thermal histories. If the dark sector states thermalise among each other but not with the SM thermal bath, one can define a dark sector temperature $T_D$ that may differ from the SM one~\cite{Breitbach:2018ddu}. Such a dark sector may be populated only gradually and undergo a phase transition that does not directly affect the SM plasma. In fact, it may even be possible to have symmetry-restoring \textit{inverse} phase transitions, which have recently been studied in the context of reheating after inflation or energy injection into dark sectors~\cite{Senaha:2020mop,Ramazanov:2021eya,Buen-Abad:2023hex,Dent:2024bhi,Sui:2025szm,Barni_2026}. These studies have focused on temperature-driven transitions, assuming chemical equilibrium within the dark sector.

The focus of the present work is to investigate how chemical equilibration affects the effective potential of a dark sector and how it may trigger an inverse phase transition.
For this purpose we consider the freeze-in production and subsequent evolution of a dark sector. While the freeze-in mechanism was first proposed as a way to produce dark matter~\cite{Hall_2010}, the mechanism applies more broadly to any dark sector particles that can be produced gradually through rare decays or annihilations~\cite{Bernal:2017kxu}. In the presence of scattering processes among the produced dark sector states, their momentum distribution is driven towards a thermal shape. While kinetic equilibrium may be established rather quickly, chemical equilibrium is much harder to achieve. This is because during freeze-in production, the dark sector fluid is initially underoccupied at the phase-space level until self-number-changing reactions become efficient. Such a dark sector is then characterised not only by its temperature $T_D$ but also by a non-vanishing chemical potential $\mu$ or, equivalently, the fugacity $z = e^{\mu/T_D}$. This observation turns out to be highly relevant when considering the phase structure of the dark sector.

After the main freeze-in production has occurred, the dark sector temperature typically decreases monotonically due to redshifting. The naive expectation would therefore be that only ordinary (symmetry-breaking) phase transitions may occur.  However, we show that the thermal corrections to the effective potential depend on both the dark sector temperature and the fugacity independently. As soon as self-number-changing reactions become efficient, they turn a small population of energetic particles into one with lower  velocities but increased phase-space occupation~\cite{Cervantes_2024,Bernal:2015xba, Bernal:2025osg, Cervantes:2026cfq}. As a result of this chemical equilibration, the fugacity grows rapidly, which enhances the thermal corrections to the effective potential and potentially restores the symmetry, even though the dark sector continues to cool. To the best of our knowledge, this is the first study demonstrating how chemical equilibration can drive symmetry restoration.

\begin{figure}[t]
    \centering
    \includegraphics[width=\textwidth]
    {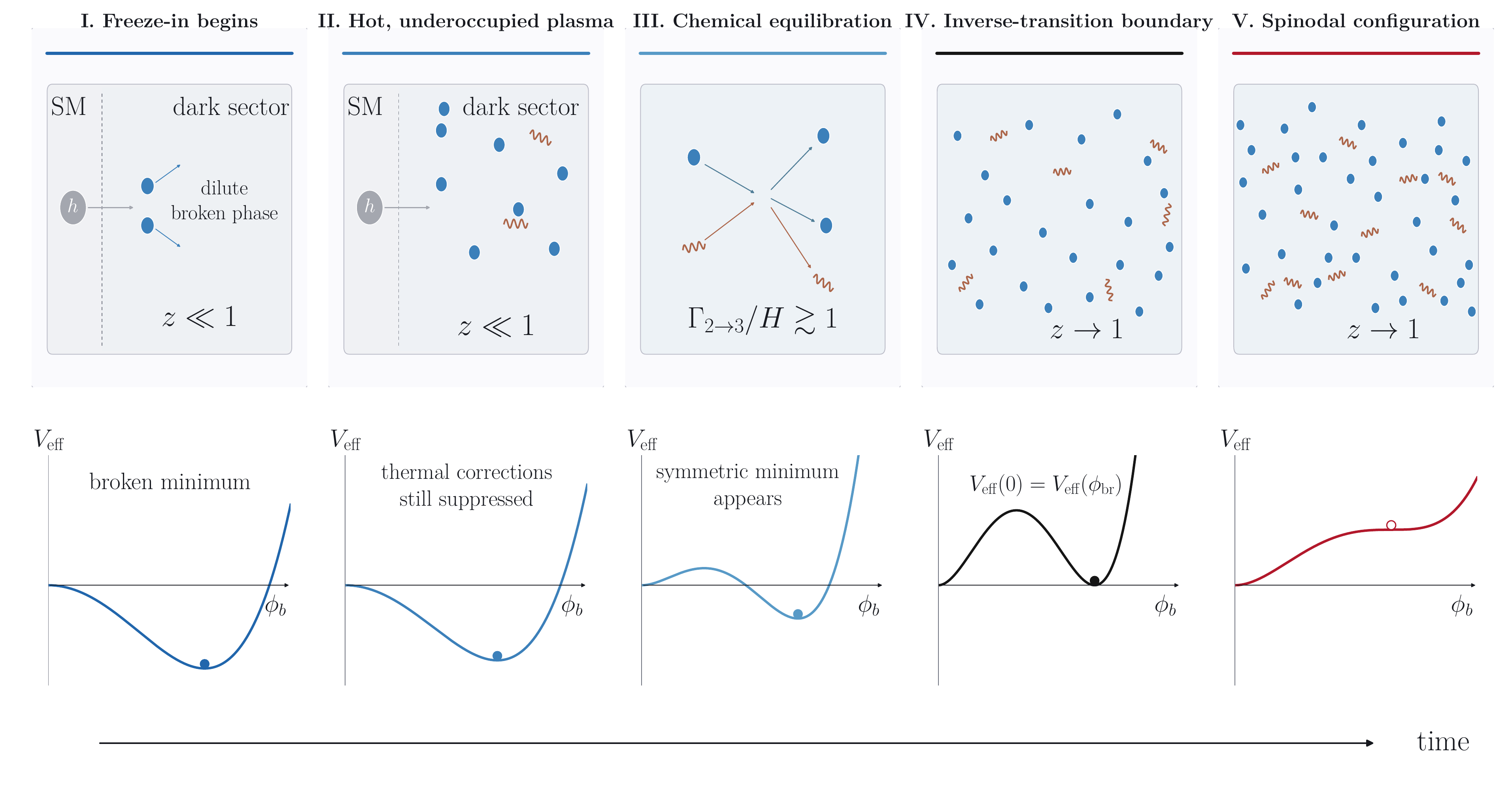}
    \caption{Schematic evolution of the dark sector during freeze-in. The upper row depicts the initial population through Higgs decays, the formation of a hot but underoccupied plasma, the onset of efficient number-changing reactions (assisted by the dark gauge boson), and the approach to chemical equilibrium. The lower row illustrates the associated evolution of the effective potential.
    }
    \label{fig:chemical_equilibration_schematic}
\end{figure}

Concretely, the dark sector that we consider consists of a new $U(1)'$ gauge symmetry and a dark Higgs field, which spontaneously breaks the gauge symmetry through its vacuum expectation value. Such a dark sector can be populated either via the Higgs portal or via the kinetic mixing portal, and we focus on the former case for simplicity. It is well-known that the interactions between the new gauge boson and the dark Higgs field create a potential barrier through thermal corrections~\cite{Quiros:1999jp, Curtin:2016urg}. Interestingly, those same interactions can lead to efficient self-number-changing processes involving dark vector and dark Higgs bosons. In other words, the interactions that generate a barrier in the effective potential are also responsible for efficiently creating the particle population needed for thermal symmetry restoration. We study this effect by solving the temperature and number density Boltzmann equations, tracking the phase of the effective potential throughout the evolution. An example of this evolution is sketched in figure~\ref{fig:chemical_equilibration_schematic}.

While we clearly establish that an inverse phase transition may occur in our set-up, determining the order of the phase transition is more difficult.
The dark sector that we consider does in principle provide the ingredients for a first-order phase transition, i.e.\ two minima in the effective potential separated by a potential barrier. Symmetry restoration would thus be expected to proceed through bubble nucleation and expansion~\cite{PhysRevD.15.2929,LINDE1983421}. However, we will see that chemical equilibration proceeds on timescales much shorter than the overall evolution of the system. This suggests that there may not be enough time for the nucleation rate to build up and drive percolation. If bubbles fail to form and instead the transition follows a spinodal history, the metastable broken phase would eventually lose local stability; the subsequent scalar evolution would then drive the system away from the broken configuration in a second-order phase transition. In the present work, we are not concerned with establishing the order of the transition; our aim instead is simply to demonstrate that the inverse phase transition is driven by chemical equilibration. 

The remainder of the paper is organised as follows. In section~\ref{sec:model} we introduce the dark Abelian Higgs model, its portal interaction with the SM, the reactions responsible for internal equilibration and its effective potential. In section~\ref{sec:results} we discuss the evolution of the temperature and fugacity for a benchmark point, emphasising the impact of the chemical potential on the effective potential and the approach to the thermodynamic boundary of the inverse phase transition. We also discuss cosmological bounds on this scenario in the case that the dark Higgs boson constitutes a fraction of the total dark matter energy density. We conclude with a discussion of the general implications of our results in section~\ref{sec:conclusions}.

\section{The dark sector and its phase structure}
\label{sec:model}

\subsection{The model}
\label{sec:model_lagrangian}

We extend the Standard Model by a dark sector consisting of a complex scalar $\Phi$ charged
under a dark $U(1)'$ gauge symmetry with corresponding gauge field denoted by $A'_\mu$. The
Lagrangian describing the model is~\cite{Holdom:1985ag,Ertas:2021xeh}
\begin{equation}
    \begin{aligned}
        \mathcal{L} &= \mathcal{L}_{\rm SM} -\frac{1}{4}F'_{\mu\nu}F'^{\mu\nu} +(D_\mu\Phi)^\dagger D^\mu\Phi -V(H,\Phi)\,, \end{aligned}
    \label{eq:model_lagrangian}
\end{equation}
where the covariant derivative is given by $D_\mu = \partial_\mu-igA'_\mu$ and the dark photon field
strength tensor is denoted by $F'_{\mu\nu}=\partial_\mu A'_\nu-\partial_\nu A'_\mu$. The tree-level potential of the
SM and dark Higgs is\footnote{The factor $1/6$ in the quartic term for the dark scalar is
  chosen to match the normalization $1/4!$ after expanding around the VEV.}
\begin{align}\label{eq:model_potential}
  V(H,\Phi) &= -\mu_h^2 (H^\dagger H) +\lambda_h(H^\dagger H)^2 -\mu_\Phi^2(\Phi^\dagger\Phi) +\frac{\lambda}{6}(\Phi^\dagger\Phi)^2 + \lambda_{h\phi}(H^\dagger H)(\Phi^\dagger\Phi)\,,
\end{align}
with the portal coupling $\lambda_{h\phi}$. At zero temperature both symmetries are broken and
hence the fields can be expanded around their expectation values $v_{h}$ and $v_{\phi}$
\begin{equation}
  H = \frac{1}{\sqrt{2}}
  \begin{pmatrix}
        0\\
        v_h+h
    \end{pmatrix}\,,
    \qquad \Phi =\frac{1}{\sqrt{2}}\left(v_\phi+\phi+i\varphi\right)\,.
    \label{eq:field_expansion}
\end{equation}
Here we write the SM Higgs field in unitary gauge and decompose $\Phi$ into the radial dark Higgs
excitation $\phi$ and the Goldstone mode $\varphi$. In the feeble-portal limit the mixing between
$h$ and $\phi$ is negligible and to leading order in $\lambda_{h\phi}$, the dark-sector masses in
vacuum are $m_\phi^2 = \lambda v_\phi^2/3$ and $m_{A'}^2 = g^2v_\phi^2$. Hence, we use
$v_\phi$, $\lambda$, $g$, and $\lambda_{h\phi}$ as the independent parameters. In the present work we fix $v_\phi = 100 \, \mathrm{MeV}$, such that the masses of the dark photon and the dark Higgs boson satisfy $m_{\phi}, m_{A^{\prime}} \ll m_{h}$, ensuring that the dark sector states are produced highly relativistically in decays of the SM Higgs boson.

If the mixed quartic coupling $\lambda_{h\phi}$ is small enough, the Higgs-portal interaction allows for the freeze-in production of dark sector states through decays and annihilations~\cite{Chu_2012,Yaguna:2011qn,Heeba_2018}. In the limit $m_\phi\ll m_h$ the leading production modes are on-shell decays of SM Higgs bosons, $h \to \phi \phi$, which give the main contribution when the temperature of the SM bath is approximately $T \approx m_h / 3$. The Higgs portal therefore provides a conceptually simple way to produce the dark sector at a well-defined period in the early universe that is cleanly separated from the dark sector internal dynamics responsible for chemical equilibration and the thermal barrier. 

Besides the Higgs portal coupling, another renormalisable operator is the kinetic mixing portal between the dark photon and the SM photon. For this portal interaction, freeze-in proceeds in several stages, including a phase of resonant dark photon production, which leads to an increasing dark sector temperature that may trigger an inverse phase transition (see e.g. ref.~\cite{Cervantes:2026vdg}). These additional effects are interesting but not necessary to demonstrate the mechanism discussed here. For simplicity, we therefore assume that the Higgs-portal coupling dominates the freeze-in dynamics of the dark sector.

\subsection{Effective potential}
\label{sec:finite_fugacity}

The phase structure of the model is encoded in the free energy of the system or, equivalently, in the effective potential. The zero-temperature potential of the dark Higgs field consists of the tree-level contribution, which is
\begin{equation}
  \label{eq:tree-level-potential}
V_{\rm tree}(\phi_b)=
-\frac{\lambda v_\phi^2}{12}\phi_b^2
+\frac{\lambda}{24}\phi_b^4\,,
\end{equation}
to which we add the radiative Coleman-Weinberg (CW) correction~\cite{Weinberg:1973am}. In the Landau gauge and $\overline{\mathrm{MS}}$ renormalisation scheme the CW potential has the form~\cite{Athron_2024}
\begin{equation}
  V_{\rm CW}(\phi_b) = \sum_{i=\phi,\varphi,A'} \frac{g_i m_i^4(\phi_b)}{64\pi^2}
  \left[\log\left(\frac{m_i^2(\phi_b)}{\Lambda^2}\right) -c_i \right]\,,
  \label{eq:CW_potential}
\end{equation}
where $g_i$ counts the degrees of freedom, $m_{i}(\phi_{b})$ denotes the field-dependent masses, $c_\phi=c_\varphi=3/2$ and $c_{A'}=5/6$ are the renormalisation constants, and $\Lambda$ is the renormalisation scale, which we set to $\Lambda = v_\phi$. The background field dependent masses are
\begin{equation}
m_\phi^2(\phi_b) = \frac{\lambda}{6}\left(3\phi_b^2-v_\phi^2\right)\,,
\qquad
m_\varphi^2(\phi_b) = \frac{\lambda}{6}\left(\phi_b^2-v_\phi^2\right)\,, \qquad
m_{A'}^2(\phi_b) = g^2\phi_b^2\,.
\label{eq:field_dependent_masses}
\end{equation}
Additionally, we also add the counter-term potential $V_{\rm ct}=-\delta\mu^2\phi_b^2/2+\delta\lambda\phi_b^4/24$, whose parameters are fixed by requiring that the zero-temperature vacuum expectation value and dark-Higgs mass retain their tree-level values:
\begin{equation}
\left.
\frac{\partial(V_{\rm CW}+V_{\rm ct})}{\partial\phi_b}
\right|_{\phi_b=v_\phi}
=
0,
\qquad
\left.
\frac{\partial^2(V_{\rm CW}+V_{\rm ct})}{\partial\phi_b^2}
\right|_{\phi_b=v_\phi}
=
0.
\label{eq:renormalization_conditions}
\end{equation}

As the freeze-in through the Higgs portal leads to the formation of a thermal bath, the effective potential receives finite-temperature corrections. Although initially only $\phi$ particles are produced, dark photons can be created through conversion and scattering processes within the dark sector. We assume that these processes are efficient and describe all dark states as a single fluid with a common temperature $T_D$ and chemical potential $\mu$.\footnote{To test local self-thermalization, one would in principle need to solve the Boltzmann equation at the phase-space level using tools such as \texttt{BEST}~\cite{Yoon:2026rce} or \texttt{KineticXGPU}~\cite{Cervantes:2026fbk}. However, the SM Higgs bosons are still semi-relativistic when they decay, and the injected dark sector particles inherit momenta similar to those of the SM thermal bath, even if self-scatterings are not efficient at early stages~\cite{Cervantes:2026fbk}. Thus, a thermal ansatz remains well justified.} Their phase-space distributions can then be parametrised as
\begin{equation}
f_i(p)
=\left[
z^{-1}\exp\left(\frac{E_i}{T_D}\right) -\eta_i \right]^{-1},
\qquad
z = \exp\left(\frac{\mu}{T_D}\right),
\label{eq:common_distribution}
\end{equation}
where the fugacity $z$ encodes the departure from chemical equilibrium~\cite{Calzetta:2008iqa}, $i$ labels the dark states, and $\eta_i=+1$ ($-1$) for bosons (fermions).

In chemical equilibrium, the thermal effects are commonly computed in the imaginary-time (Matsubara) formalism~\cite{Quiros:1999jp,Athron_2024,Kapusta:2006pm}. Here the out-of-chemical-equilibrium scenario requires instead adopting the real-time quasiparticle description~\cite{Lundberg:2020mwu,Bellac:2011kqa, LANDSMAN1987141}, within which the contribution to the effective potential corresponds to the negative pressure of the dark-sector species and can be expressed in terms of their phase-space distributions
\begin{equation}
\begin{aligned}
V_{\rm med}(\phi_b;T_D,z) &\equiv-P_D= -\sum_i g_i \int_p \frac{p^2}{3E_i}\, f_i(p) \\
                       &= \frac{T_D^4}{2\pi^2} \sum_i \frac{g_i}{\eta_i}\int_0^\infty dy\,y^2\log\left[1-\eta_i z\exp\left(-\sqrt{y^2+x_i^2}\right) \right]\,,
\end{aligned}
\label{eq:medium_potential}
\end{equation}
where $\int_p\equiv \int\frac{d^3p}{(2\pi)^3}$, $E_i=\sqrt{p^2+m_i^2(\phi_b)}$ and
$x_i=m_i(\phi_b)/T_D$. The usual equilibrium result is recovered for $z = 1$. While there is
no closed form for eq.~\eqref{eq:medium_potential}, a useful semi-analytic representation
is obtained by expanding the distribution function as
\begin{equation}\label{eq:dist_expansion}
	f_i
	=
	\frac{1}{z^{-1}e^{E_i/T_D}-\eta_i}
	=
	\frac{z\,e^{-E_i/T_D}}
	{1-\eta_i z\,e^{-E_i/T_D}}
	=
	\sum_{k=1}^{\infty}
	\eta_i^{\,k-1}z^k e^{-kE_i/T_D}.
\end{equation}

This series converges pointwise whenever \(z\,e^{-E_i/T_D}<1\). For massive species (\(E_i\geq m_i\)) and \(0<z\leq1\), this condition is automatically satisfied for all momenta. During freeze-in, and in the absence of self-number-changing reactions, one always has \(z<1\), such that the series expansion is well defined. Even when self-number-changing reactions later set \(z=1\), the expansion remains pointwise convergent for massive species, although in the bosonic case convergence can become slow when \(m_i/T_D\ll1\) due to Bose enhancement.

Using the expansion in eq.~(\ref{eq:dist_expansion}), we compute the leading order contribution as follows:
\begin{equation}
\begin{aligned}
  \mathcal{V}_{i}\left( m^2_{i}; T_{D}, z \right)
  = - P_i(m_i^2;T_D,z)
  =
  - g_i\,\frac{m_i^2T_D^2}{2\pi^2}\sum_{k=1}^{\infty}\eta_i^{\,k-1}\frac{z^k}{k^2}K_2\left(\frac{k m_i}{T_D}\right)\,,
\end{aligned}
\label{eq:species_pressure}
\end{equation}
where $g_i$ denotes the number of degrees of freedom associated with the species under consideration. The total unscreened medium contribution is $V_{\rm med}=\sum_i\mathcal V_i$. The expressions above apply to both bosonic and fermionic species. The dark-sector spectrum considered in the present work consists entirely of bosons; thus, we set $\eta_i=1$ for all species in what follows.

The series converges rapidly in the underoccupied regime relevant for the phase  transition. Although smaller cutoffs already provide sufficient numerical accuracy, we conservatively adopt the fixed value $k_{\max}=8$. The remainder of the geometric expansion is bounded by $z^{k_{\max}} e^{-k_{\max} E_i/T_D}$, and is therefore suppressed both by powers of the fugacity and by the Boltzmann factor. This estimate is moreover conservative for the pressure in eq.~\eqref{eq:species_pressure}, since the modified Bessel function decreases with increasing $k$, while each term carries an additional $1/k^2$ suppression. In appendix~\ref{app:kf_convergence} we consider an example benchmark point with $z \simeq 0.02$, for which we find that $k_{\max}=8$ guarantees the truncation error to be numerically negligible.

\subsection{Screened thermal potential}
\label{sec:screened_V}

Let us now turn to the issue of resummation. Soft bosonic modes require a reorganization of perturbation theory when medium-induced self-energies become comparable to their field-dependent squared masses~\cite{Linde:1978px,Gross:1980br}. Nonetheless, away from chemical equilibrium this resummation cannot be obtained by inserting a self-energy into the ring potential, as performed in chemical equilibrium~\cite{Parwani:1991gq,Arnold:1992rz}. Instead, in the real-time formalism soft retarded propagators can be dressed by self-energies determined by the actual phase-space distributions~\cite{Carrington:1997sq}. Related resummations have been formulated in terms of distribution-dependent tadpoles and self-consistent propagators in ref.~\cite{Riotto:1996vi}. In equilibrium scalar theories, dressing the propagator in the one-loop tadpole provides the partial-dressing construction and reproduces the relevant daisy and superdaisy combinatorics~\cite{Boyd:1993tz}.

For the $U(1)'$ model, the contributions to the one-loop static self-energies are
\begin{equation}
\begin{aligned}
\Pi_\phi
&=
\frac{\lambda}{2}\,\mathcal I_\phi
+\frac{\lambda}{6}\,\mathcal I_\varphi
+3g^2\,\mathcal I_{A'},
\\
\Pi_\varphi
&=
\frac{\lambda}{6}\,\mathcal I_\phi
+\frac{\lambda}{2}\,\mathcal I_\varphi
+3g^2\,\mathcal I_{A'},
\\
\Pi_{A'_L}
&=
2g^2\left(\mathcal I_\phi+\mathcal I_\varphi\right)\,,
\end{aligned}
\label{eq:thermal_self_energies}
\end{equation}
where the tadpole integral for a bosonic species is
\begin{equation}
\begin{aligned}
\mathcal I_i(\phi_b;T_D,z)
=\int_p\frac{f_i(p)}{E_i(\phi_b)}
=
\frac{m_i(\phi_b)T_D}{2\pi^2}
\sum_{k=1}^{\infty}
\frac{z^k}{k}
K_1\left(
\frac{k m_i(\phi_b)}{T_D}
\right).
\end{aligned}
\label{eq:tadpole_integral}
\end{equation}
The derivation of these formulae can be found in appendix~\ref{app:finite_z_screening}. In the relativistic limit this integral becomes $\mathcal I_i(T_D,z) = T_D^2/(2\pi^2)\operatorname{Li}_2(z)$, where \(\operatorname{Li}_2(z)\) denotes the dilogarithm (polylogarithm of order two), yielding $T_D^2/12$ in chemical equilibrium. To construct the screening self-energies, we take this relativistic limit, which is a suitable approximation for phase transitions where the scalar modes are nearly massless and the gauge mode becomes increasingly soft as the symmetry-restoring region is approached. This choice also has the useful consequence that, at fixed $(T_D,z)$, the leading self-energies are independent of the background field. 

The medium function satisfies
\begin{equation}
\frac{\partial\mathcal V_i}{\partial m_i^2}
=
\frac{g_i}{2}\mathcal I(m_i;T_D,z).
\label{eq:medium_tadpole_identity}
\end{equation}
Dressing the propagator therefore amounts to replacing $m^2\rightarrow M^2=m^2+\Pi$ in the medium function. The corresponding screened masses are
\begin{equation}
\begin{aligned}
M_\phi^2(\phi_b;T_D,z)
&=
m_\phi^2(\phi_b)+\Pi_\phi(T_D,z)\,,
\\
M_\varphi^2(\phi_b;T_D,z)
&=
m_\varphi^2(\phi_b)+\Pi_\varphi(T_D,z)\,,
\\
M_{A'_L}^2(\phi_b;T_D,z)
&=
m_{A'}^2(\phi_b)+\Pi_{A'_L}(T_D,z)\,,
\end{aligned}
\label{eq:screened_masses}
\end{equation}
while the transverse gauge modes are not screened at this order. In the relativistic approximation, $\Pi_i$ depends on $T_D$ and $z$ but not on the background field, so this mass replacement is equivalent to integrating the partially dressed tadpole up to a $\phi_b$-independent contribution. Further details are given in appendix~\ref{app:finite_z_screening}. 

Putting everything together, we obtain the explicit form of the screened medium contribution to the effective potential:
\begin{equation}
\begin{aligned}
V_{\rm med}^{\rm scr}(\phi_b;T_D,z)
\equiv{}&
\mathcal V_\phi(M_\phi^2;T_D,z)
+
\mathcal V_\varphi(M_\varphi^2;T_D,z)
\\
&+
\,\mathcal V_{A'_T}(m_{A'_T}^2;T_D,z)
+
\mathcal V_{A'_L}(M_{A'_L}^2;T_D,z).
\end{aligned}
\label{eq:screened_medium_potential}
\end{equation}
For the screened medium contribution it is necessary to split the dark photon into transverse and longitudinal polarizations via
\begin{equation}
g_\phi=g_\varphi=1,
\qquad
g_{A'_T}=2,
\qquad
g_{A'_L}=1.
\label{eq:screened_degrees_of_freedom}
\end{equation}
In total, the effective potential consists of the terms
\begin{equation}
V_{\rm eff}(\phi_b;T_D,z)=
V_{\rm tree}(\phi_b)
+V_{\rm CW}(\phi_b)
+V_{\rm ct}(\phi_b)
+V_{\rm med}^{\rm scr}(\phi_b;T_D,z)\,.
\label{eq:effective_potential}
\end{equation}
Throughout this work, the loop and screened medium contributions are evaluated in Landau gauge. The Goldstone mode is therefore retained as part of the gauge-fixed loop expansion, as in standard treatments of the finite-temperature effective potential~\cite{Wainwright:2011qy,Garny:2012cg,Athron_2024,Croon:2020cgk,Kapusta:2006pm}. As usual for finite-order effective-potential calculations, the resulting phase structure retains a residual gauge dependence; alternative gauge formulations and the associated perturbative subtleties are discussed in refs.~\cite{Tye:1996au,Irges:2017ztc,Arnold:1992fb,Arnold:1992cg,Chaichian:1991aa,NIELSEN1975173,Patel:2011th}.

\subsection{Phase structure and transition boundaries}
\label{sec:phase_structure}

To define a thermodynamic boundary of the phase transition we require the coexistence of a symmetric and a broken minimum with $\phi_{\rm br}\neq\phi_{\rm sym}$, which both satisfy
\begin{equation}
\frac{\partial V_{\rm eff}\bigl(\phi_j;T_D,z\bigr)}{\partial\phi_b}
=
0,
\qquad
\frac{\partial^2 V_{\rm eff}\bigl(\phi_j;T_D,z\bigr)}{\partial\phi_b^2}
>
0,
\qquad
j\in\{\mathrm{sym},\mathrm{br}\}.
\label{eq:inverse_transition_minima}
\end{equation}
For the field convention adopted here, $\phi_{\rm sym}=0$, whereas $\phi_{\rm br}\neq0$. 
The thermodynamic degeneracy boundary is then defined as the set of dark sector temperatures and fugacities for which the symmetric and broken minima have equal free energy:
\begin{equation}
V_{\rm eff}\bigl(\phi_{\rm br};T_D,z\bigr)
=
V_{\rm eff}\bigl(\phi_{\rm sym};T_D,z\bigr) \; .
\label{eq:inverse_transition_criterion}
\end{equation}
For temperatures and fugacities below this boundary, the broken minimum corresponds to the global one, while above the boundary the symmetric minimum has smaller free energy and the broken minimum becomes metastable. Further away from the boundary, the metastable minimum eventually disappears completely and only one minimum remains. The point where the broken minimum loses local stability, called the spinodal point, is defined by
\begin{equation}
\frac{\partial V_{\rm eff}\bigl(\phi_{\rm spinodal};T_D,z\bigr)}{\partial\phi_b}
=
0,
\qquad
\frac{\partial^2 V_{\rm eff}\bigl(\phi_{\rm spinodal};T_D,z\bigr)}{\partial\phi_b^2}
=
0.
\label{eq:broken_spinodal}
\end{equation}
The coexistence of two locally stable phases and the subsequent loss of metastability at the spinodal are standard features of first-order phase transitions~\cite{K_Binder_1987}.

\begin{figure}[t]
    \centering
    \includegraphics[width=0.9\textwidth]{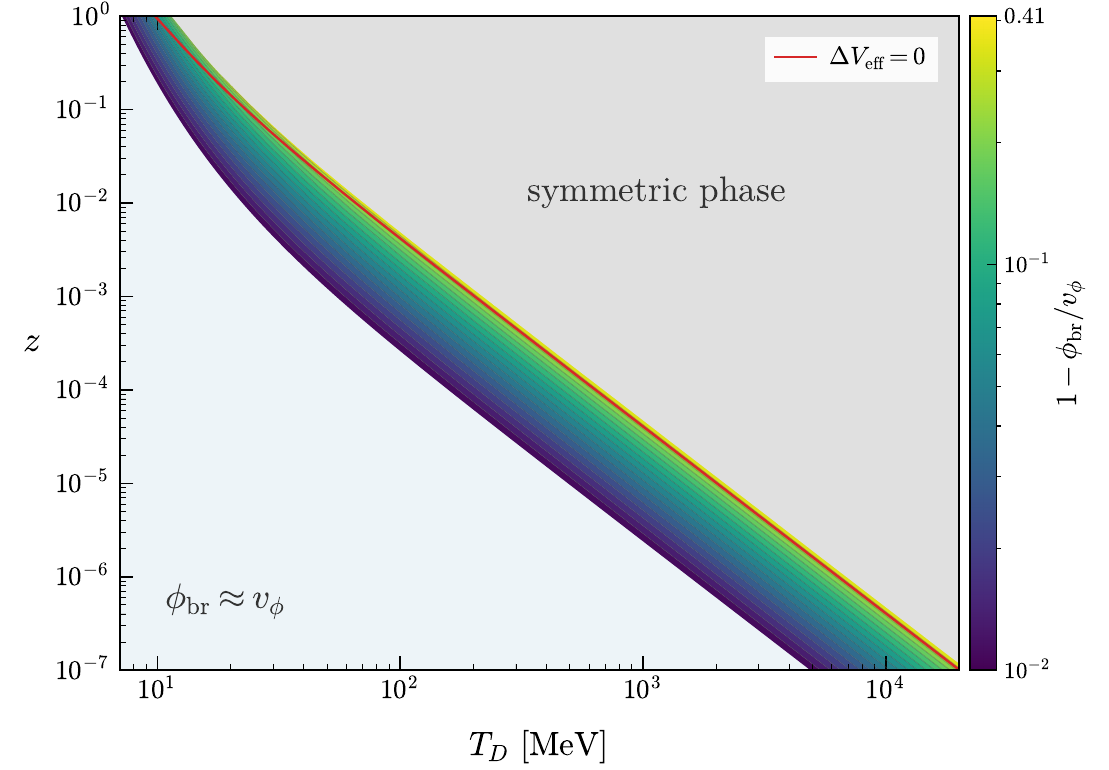}

    \caption{ Phase structure in the $(T_D,z)$ plane for the benchmark point in table~\ref{tab:benchmark}. The light blue--green area highlights the vacuum-like broken phase below the displayed cutoff $1-\phi_{\rm br}/v_\phi=10^{-2}$, while the light-gray area highlights the symmetric phase. The red solid curve marks broken--symmetric degeneracy, $\Delta V_{\rm eff} \equiv V_{\rm eff}(\phi_{\rm br})-V_{\rm eff}(0)=0$.}
  \label{fig:vev_scan_perturbativity}
\end{figure}
%

The resulting phase diagram is shown in figure~\ref{fig:vev_scan_perturbativity} for $g = 0.3$ and $\lambda = 10^{-3}$.  The hierarchy $\lambda \ll g$ of the couplings ensures that the potential develops a barrier. In addition to the degeneracy boundary (red line), we also show the field value at the broken minimum, $\phi_{\rm br}$, as a function of $T_{D}$ and $z$ up to the spinodal point.

For fixed
fugacity, increasing $T_D$ progressively displaces the broken minimum toward the
symmetric configuration, while increasing the occupation at fixed $T_D$ has an analogous
restoring effect. In the dilute regime ($z \ll 1$), $\operatorname{Li}_2(z)\simeq z$, so the
leading restoring self-energies scale approximately as $zT_D^2$. In this regime the degeneracy boundary therefore scales as $z \propto T_D^{-2}$. The central conclusion is that it is possible for the dark sector to cross from the broken into the symmetric phase without increasing its temperature, provided that there is a rapid increase in fugacity. In the following section we will show that this is exactly what happens during chemical equilibration.

\begin{table}[t]
  \renewcommand{\arraystretch}{1.2}
  \centering
\begin{tabular}{ccccc}
\toprule
\(g\)    & \(\lambda\) & \(v_{\phi}\) & \(m_{A^{\prime}}\) & \(m_{\phi}\) \\ [0.5ex] \midrule
\(0.3\)  & \(10^{-3}\) & $100\,$MeV & $30\,$MeV & $1.83\,$MeV \\
\bottomrule
\end{tabular}
\caption{Model parameters and particle masses of our benchmark point.}
\label{tab:benchmark}
\end{table}

\section{Chemical equilibration and the inverse transition}
\label{sec:results}

Having discussed the phase structure of the dark sector, we now turn to the evolution of the dark sector temperature and fugacity. We focus on the same benchmark point as above, which is summarized in table~\ref{tab:benchmark}. 

At the beginning of the freeze-in process, decays of the SM Higgs boson inject particles and energy
into the dark sector, which initially is in the broken phase. While the dark sector
remains dilute ($z \ll 1$), the thermal and medium effects on the potential are suppressed,
keeping the dark sector in the broken phase. Once the $2\to3$ reactions become efficient,
they drive the system towards chemical equilibrium. The resulting growth in occupancy
enhances the thermal corrections to the effective potential, leading to the appearance of
a symmetric minimum and a potential barrier. The description of this process requires a
careful treatment of the temperature and fugacity equations.

\subsection{Temperature and fugacity equations}
\label{sec:thermodynamic_closure}

The conventional way to study the evolution of the dark sector temperature and fugacity would be to study moments of the Boltzmann equations for $\phi$ and $A'$, assuming the dynamics of the background
field $\phi_b$ to be negligible. In the scenario considered here, however, chemical
equilibration modifies the medium corrections to the effective potential and progressively
displaces the broken minimum, so that the feedback of the evolving background on the
dark-sector energy density and quasiparticle spectrum must be taken into account. The
vacuum and medium contributions to the dark-sector free energy are encoded in the effective potential, while the corresponding background-field and medium-dependent quasiparticle masses enter the dispersion relations. A possible way to systematically treat the coupled background and particle dynamics from first principles could be the closed-time-path formalism, using a two-particle-irreducible
effective action to evolve the background field and particle correlators simultaneously
(see e.g.~refs.~\cite{Calzetta:1986cq,Ai:2023qnr,Becker:2023vwd} and references therein).
However, given the separation between the fast microphysical relaxation of $\phi_b$ and the much slower freeze-in evolution, we instead adopt a simpler approach based on an adiabatic effective-potential closure.

We formulate the energy evolution directly in terms of the dark-sector
energy-momentum tensor,
\begin{equation}
\nabla_\mu T_D^{\mu\nu}=
J^\nu,
\label{eq:energy_momentum_source}
\end{equation}
where $J^\nu$ describes the transfer of energy and momentum from the Standard Model to the dark sector. For a homogeneous and isotropic fluid with no net momentum injection,
\begin{equation}
J^\nu
=
\left(
C_E^{\rm FI},
0,
0,
0
\right)\,,
\label{eq:dark_sector_source}
\end{equation}
where $C_E^{\rm FI}$ denotes the energy-transfer rate from the Standard Model to the dark
sector. In a kinetic description, it corresponds to the energy-weighted moment of the
freeze-in collision operator, which we can calculate explicitly for a given source. The temporal component of eq.~\eqref{eq:energy_momentum_source}
then gives
\begin{equation}
\dot{\rho}_D + 3H\left( \rho_D+P_D \right)
=
C_E^{\rm FI}\,.
\label{eq:energy_density_evolution}
\end{equation}

The pressure and energy density are obtained from the effective potential. At each time,
we assume that the homogeneous background field $\phi_{b}$ follows the relevant local
minimum of the effective potential adiabatically
\begin{equation}
\left.
\frac{\partial V_{\rm eff}(\phi_b;T_D,z)}{\partial\phi_b}
\right|_{\phi_b=\phi_{\min}(T_D,z)}
=
0\,.
\label{eq:instantaneous_minimum}
\end{equation}
Then the pressure and energy density in this minimum are given by
\begin{align}
P_D(T_D,z)
&=
-\tilde V(T_D,z) \,,
\\
\rho_D(T_D,z)
&=
\tilde V(T_D,z)
-
T_D
  \frac{\partial\tilde V(T_D,z)}{\partial T_D}\,,
\label{eq:thermodynamics_from_reduced_potential}
\end{align}
where we denoted the effective potential
evaluated at the broken phase as
\begin{equation}
\tilde V(T_D,z)
\equiv
V_{\rm eff}\!\left(
\phi_{\min}(T_D,z);T_D,z
\right)\,.
\label{eq:reduced_effective_potential}
\end{equation}
On the other hand, the total number density is
\begin{equation}
n_D(T_D,z)
=
\sum_i\,
n_i\!\left(T_D,z,M_i^2(T_D,z)\right)\,,
\qquad
i\in\{\phi,A'_T,A'_L\}\,,
\label{eq:physical_quasiparticle_number}
\end{equation}
where the transverse component of $A^{\prime}$ has $g_{A'_T}=2$, while $\phi$ and the longitudinal
component of $A^{\prime}$ each have one degree of freedom. The total number density evolves as
\begin{equation}
\dot{n}_D
+
3Hn_D
=
C_0^{\rm FI}
+
C_0^{\rm nc}\,,
\label{eq:number_density_evolution}
\end{equation}
where $C_0^{\rm FI}$ describes the injection of particles through the Higgs portal, while
$C_0^{\rm nc}$ contains the internal number-changing reactions, which conserve the total
dark-sector energy and do not provide an additional source term in
eq.~\eqref{eq:energy_density_evolution}. 
The freeze-in source terms arise from decays of the SM Higgs boson into dark Higgs particles.
Integrating over the equilibrium SM Higgs boson distribution gives
\begin{align}
C_0^{\rm FI}&=\frac{\lambda_{h\phi}^2v_h^2m_h T_{\rm SM}}{16\pi^3}\sqrt{1-\frac{4m_\phi^2}{m_h^2}} K_1\left( \frac{m_h}{T_{\rm SM}} \right)
, \label{eq:higgs_decay_C0}
\\
C_E^{\rm FI}
&=\frac{\lambda_{h\phi}^2v_h^2m_h^2T_{\rm SM}}{32\pi^3}\sqrt{1-\frac{4m_\phi^2}{m_h^2}}K_2\left(\frac{m_h}{T_{\rm SM}}\right),
\label{eq:higgs_decay_CE}
\end{align}
where $K_n$ is the modified Bessel function of the second kind and order $n$, $T_{\rm SM}$
is the temperature of the Standard Model bath, $m_h$ is the SM Higgs-boson mass, and $v_h$
the vacuum expectation value of the Higgs field. For $m_\phi\ll m_h$, the production from on-shell Higgs decays
dominates, and we can neglect annihilation processes~\cite{Heeba_2018}.

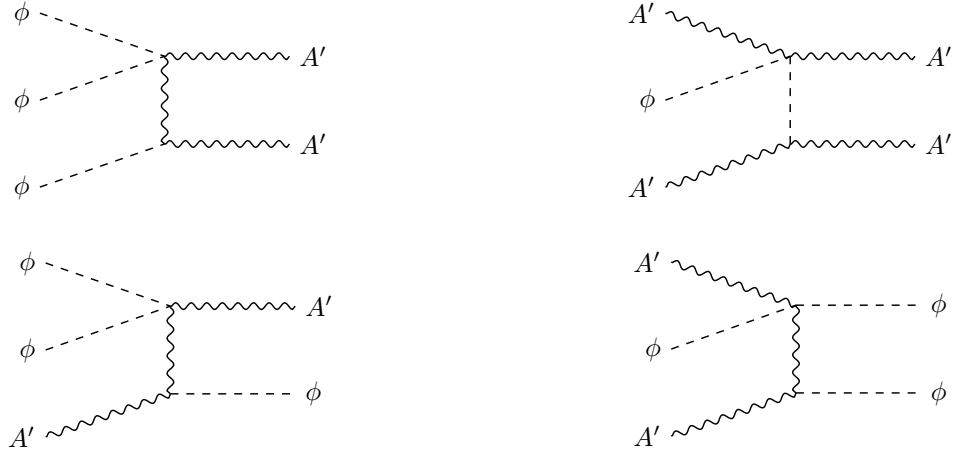
\begin{figure}[t]
  \centering

  \begin{minipage}[t]{0.47\textwidth}
    \centering
    \begin{tikzpicture}[
      baseline=(current bounding box.center),
      every label/.style={font=\small}
    ]
      \cannibalvertices{\phi}{\phi}{\phi}{A'}{A'}

      \darkscalarline{i1}{u}
      \darkscalarline{i2}{u}
      \darkphotonline{u}{o1}
      \darkphotonline{u}{d}
      \darkscalarline{i3}{d}
      \darkphotonline{d}{o2}
    \end{tikzpicture}
  \end{minipage}
  \hfill
  \begin{minipage}[t]{0.47\textwidth}
    \centering
    \begin{tikzpicture}[
      baseline=(current bounding box.center),
      every label/.style={font=\small}
    ]
      \cannibalvertices{A'}{\phi}{A'}{A'}{A'}

      \darkphotonline{i1}{u}
      \darkscalarline{i2}{u}
      \darkphotonline{u}{o1}
      \darkscalarline{u}{d}
      \darkphotonline{i3}{d}
      \darkphotonline{d}{o2}
    \end{tikzpicture}
  \end{minipage}

  \vspace{0.4cm}

  \begin{minipage}[t]{0.47\textwidth}
    \centering
    \begin{tikzpicture}[
      baseline=(current bounding box.center),
      every label/.style={font=\small}
    ]
      \cannibalvertices{\phi}{\phi}{A'}{A'}{\phi}

      \darkscalarline{i1}{u}
      \darkscalarline{i2}{u}
      \darkphotonline{u}{o1}
      \darkphotonline{u}{d}
      \darkphotonline{i3}{d}
      \darkscalarline{d}{o2}
    \end{tikzpicture}
  \end{minipage}
  \hfill
  \begin{minipage}[t]{0.47\textwidth}
    \centering
    \begin{tikzpicture}[
      baseline=(current bounding box.center),
      every label/.style={font=\small}
    ]
      \cannibalvertices{A'}{\phi}{A'}{\phi}{\phi}

      \darkphotonline{i1}{u}
      \darkscalarline{i2}{u}
      \darkscalarline{u}{o1}
      \darkphotonline{u}{d}
      \darkphotonline{i3}{d}
      \darkscalarline{d}{o2}
    \end{tikzpicture}
  \end{minipage}

  \caption{Example of gauge tree-level contributions to the broken-phase $2\leftrightarrow3$ reactions in eq.~\eqref{eq:number_changing_channels}. Dashed and wavy lines denote $\phi$ and $A'$, respectively.
  }
  \label{fig:gauge_assisted_number_changing_diagrams}
\end{figure}

In the broken phase the chemical equilibration is controlled via $2\leftrightarrow3$ processes involving the dark Higgs and dark photon,
\begin{equation}
\begin{aligned}
\mathcal R_{2\leftrightarrow3}
=
\bigl\{
&2\phi\leftrightarrow3\phi,\,
2A'\leftrightarrow3\phi,\,
2A'\leftrightarrow2A'\phi,\,
\phi A'\leftrightarrow2\phi A',\,
2\phi\leftrightarrow2A'\phi
\bigr\}.
\end{aligned}
\label{eq:number_changing_channels}
\end{equation}
The number-changing-reaction collision term can be expressed as
\begin{equation}
C_0^{\rm nc}
=
\sum_{r\in\mathcal R_{2\leftrightarrow3}}
\gamma_r^{\rm eq}(T_D)
\left(z^2-z^3\right),
\label{eq:total_number_changing_source}
\end{equation}
where $\gamma_r^{\rm eq}$ contains the corresponding equilibrium reaction density, including symmetry factors. The tree-level $2\to3$ cross sections are evaluated with \texttt{CalcHEP}~\cite{Belyaev:2012qa} and thermally averaged numerically assuming a Maxwell--Boltzmann distribution. Some of the tree-level reactions are shown in figure~\ref{fig:gauge_assisted_number_changing_diagrams}. The scalar-only reaction $2\phi\leftrightarrow3\phi$ is evaluated as in ref.~\cite{Cervantes_2024}, but we find it to be negligible due to the hierarchy $\lambda\ll g$, while the dominant contributions are $2A' \to 2A'\phi$ and $A'\phi\to2\phi A'$ (cf.~appendix~\ref{app:collision_operators}). The rates $\Gamma_{2\to3}$ and $\Gamma_{3\to2}$ used in the evolution
are obtained by summing over all channels. Their construction from the cross sections is detailed in
appendix~\ref{app:collision_operators}.

To solve the differential equations derived above, it is necessary to express $T_D$ and $z$ in terms of $\rho_D$ and $n_D$. Alternatively, we can directly translate the differential equations derived above into differential equations for the dark sector temperature and fugacity as a function of the scale factor. Using  $\text{d}/\text{d}t=aH\text{d}/\text{d}a$, we rewrite the energy and number density equations as
\begin{align}
\frac{\mathrm{d}n_D}{\mathrm{d}a} & = 
-\frac{3n_D}{a}
+
\frac{C_0^{\rm FI}+C_0^{\rm nc}}{aH} \equiv \mathcal S_n \; ,
\nonumber
\\
\frac{\mathrm{d}\rho_D}{\mathrm{d}a} & =
-\frac{3(\rho_D+P_D)}{a}
+
\frac{C_E^{\rm FI}}{aH} \equiv \mathcal S_\rho \; .
\label{eq:scale_factor_sources}
\end{align}
Applying the chain rule gives
the two coupled equations
\begin{align}
\frac{\partial n_D}{\partial T_D}\frac{{\rm d}T_D}{{\rm d}a}
+
\frac{\partial n_D}{\partial z}\frac{{\rm d}z}{{\rm d}a}
&=
\mathcal S_n,
\nonumber
\\
\frac{\partial \rho_D}{\partial T_D}\frac{{\rm d}T_D}{{\rm d}a}
+
\frac{\partial \rho_D}{\partial z}\frac{{\rm d}z}{{\rm d}a}
&=
\mathcal S_\rho.
\label{eq:scale_factor_chain_rule}
\end{align}
We can simplify these expressions by defining the energy associated with
changing the particle density at fixed temperature
\begin{equation}
\varepsilon_n
\equiv
\left(\frac{\partial\rho_D}{\partial n_D}\right)_{T_D}
=
\frac{\partial \rho_D}{\partial z}\left(\frac{\partial n_D}{\partial z}\right)^{-1} \; ,
\label{eq:energy_per_particle_app}
\end{equation}
as well as the heat capacity at fixed particle density
\begin{equation}
c_n
\equiv
\left(\frac{\partial\rho_D}{\partial T_D}\right)_{n_D}
=
\frac{\partial \rho_D}{\partial T_D}
-
\varepsilon_n \frac{\partial n_D}{\partial T_D} \; .
\label{eq:fixed_number_heat_capacity_app}
\end{equation}
These quantities can be calculated directly from the effective potential as shown in appendix~\ref{app:scale_factor_evolution}.

The evolution equations can then be written in the following compact form:
\begin{align}
\frac{{\rm d}T_D}{{\rm d}a}
&=
\frac{
\mathcal S_\rho
-\varepsilon_n\mathcal S_n
}{
c_n
}\,,
\label{eq:temperature_scale_factor_evolution}
\\
\frac{{\rm d}z}{{\rm d}a}
&=
\left(\frac{\partial n_D}{\partial z}\right)^{-1}
\left[
\mathcal S_n
-
\frac{\partial n_D}{\partial T_D} \frac{{\rm d}T_D}{{\rm d}a}
\right]\,.
\label{eq:fugacity_scale_factor_evolution}
\end{align}
The first equation states that the temperature responds to the part of the energy-density evolution that is not already accounted for by the changing particle density. The second equation assigns the remaining number-density response to the fugacity. 

A final comment concerns the definition of the number density and the associated collision terms. Equation~\eqref{eq:physical_quasiparticle_number} counts the physical broken-phase quasiparticle degrees of freedom, namely the radial scalar and the three physical vector polarizations, whereas the Landau-gauge effective potential contains an explicit Goldstone contribution as part of the gauge-fixed loop expansion. Thus, we do not interpret the Goldstone contribution to $V_{\rm eff}$ as an additional particle species in the number-density equation. The collision terms are evaluated at leading order from physical broken-phase scattering and decay amplitudes and are thermally averaged within the quasiparticle approximation. Medium effects enter through the adopted quasiparticle dispersion relations. While the corresponding on-shell vacuum amplitudes are gauge independent, a fully gauge-consistent treatment of medium corrections to the collision operator would require a consistent resummation of the thermal propagators, spectral functions and vertices. Within our adiabatic closure, the resulting physical particle-number balance determines the evolution of the fugacity $z$, which is then used as the occupation parameter in the Landau-gauge medium functions. A fully unified non-equilibrium treatment of the background and quasiparticle correlators is beyond the scope of the present work.

\subsection{Dark sector evolution}
\label{sec:benchmark_evolution}

Having derived the differential equations for $T_D$ and $z$ (eqs.~\eqref{eq:temperature_scale_factor_evolution} and~\eqref{eq:fugacity_scale_factor_evolution}), we are now in a position to determine the cosmological evolution of the dark sector and establish whether an inverse phase transition can occur. 
We start by solving the coupled differential equations at $T_{\rm SM}=150\,{\rm GeV}$, which we identify with a scale factor of $a=1$. At higher temperatures, electroweak symmetry is restored, and decays of the SM Higgs boson are no longer possible. These temperatures give a negligible freeze-in contribution~\cite{Bringmann:2021sth} and can therefore be safely neglected in our analysis. Our results are insensitive to the initial conditions for $T_D$ and $z$ as long as the initial energy density and pressure are negligible compared to those injected by SM Higgs boson decay. For our numerical analysis, we set $T_D/T_{\rm SM}=3\times10^{-4}$ and $z=10^{-10}$. We then solve the coupled number- and energy-density equations for $T_D$ and $z$ for different values of $\lambda_{h\phi} \in [2.2\times10^{-15}, 5\times10^{-13}]$, while tracking the minimum of the effective potential along this evolution to determine whether the initially broken symmetry is restored.

\begin{figure}[t!]
    \centering
    \includegraphics[width=1\textwidth]
    {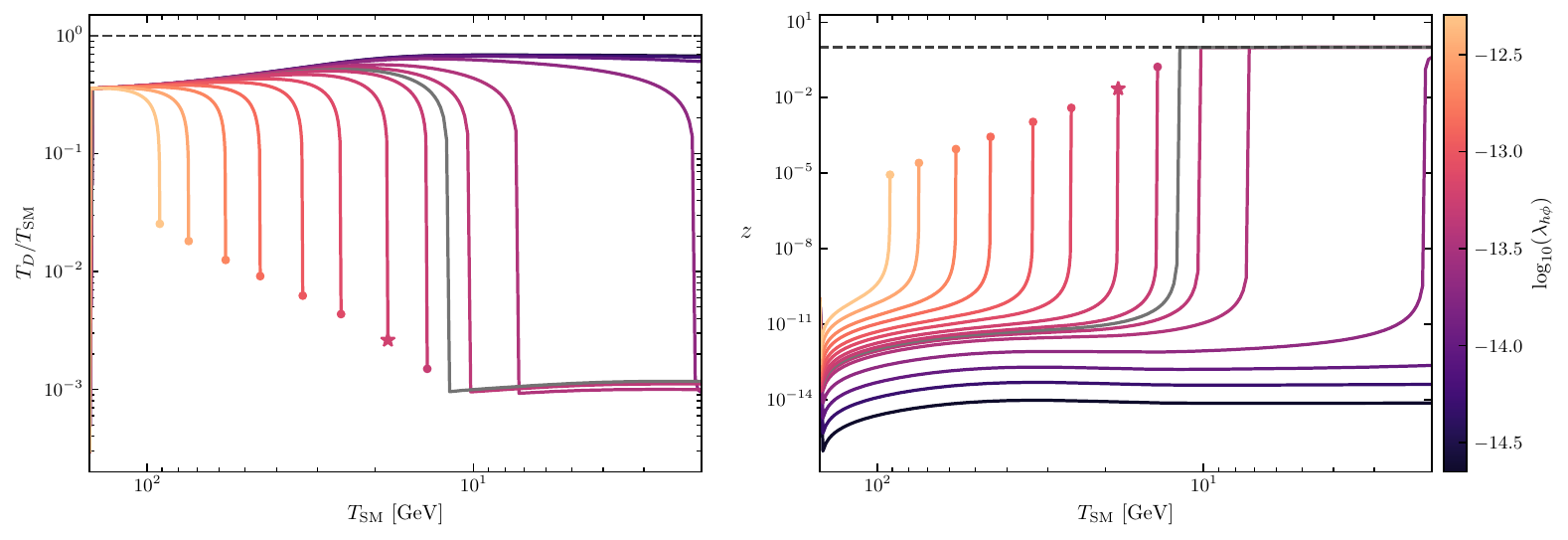}
    \caption{Evolution for the benchmark point defined in table~\ref{tab:benchmark} for different values of the Higgs portal coupling $\lambda_{h\phi}$. Left: dark-to-visible temperature ratio. Right: dark-sector fugacity. The same colour mapping for $\lambda_{h\phi}$ is used in both panels. The horizontal dashed lines indicate $T_D=T_{\rm SM}$ and $z=1$, respectively. Filled circles mark trajectories that reach the spinodal of the broken-phase evolution, while the star corresponds to the plot in figure~\ref{fig:transition_potential}. }
    \label{fig:thermodynamic_evolution}
\end{figure}

\begin{figure}[h!]
    \centering
    \includegraphics[width=0.75\textwidth]
    {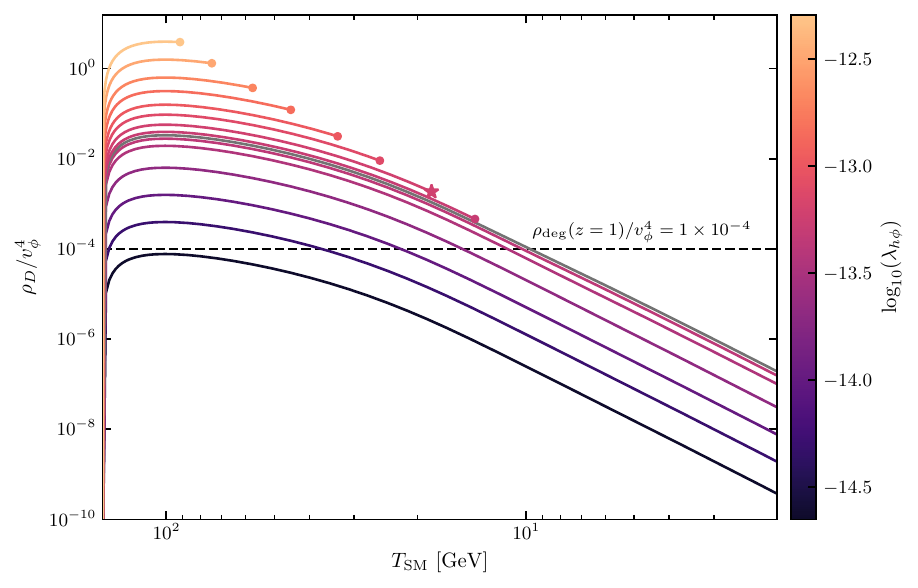}
    \caption{Dark-sector energy-density evolution for the same parameter points as in figure~\ref{fig:thermodynamic_evolution}, normalised by $v_\phi^4$. The horizontal reference line indicates $\rho_{\rm deg}/v_\phi^4=1\times10^{-4}$ (defined in the text), and filled circles mark the numerical endpoints of the broken-phase evolution.
    }
    \label{fig:energy_evolution}
\end{figure}
Figure~\ref{fig:thermodynamic_evolution} displays the resulting temperature and fugacity evolutions, which proceed from left to right as the Standard Model bath cools. At early times, the dark sector is dilute (tiny fugacity) but the temperature is comparable to that of the SM thermal bath, since the dark sector particles inherit the energy of the decaying Higgs bosons. Once the dark sector is sufficiently populated by SM Higgs boson decays, $2 \to 3$ self-interactions become efficient and drive chemical equilibration $z\to 1$. This conversion of a few high-energy particles into many lower-energy particles results in a sharp drop in the temperature (left panel) and a correspondingly sharp rise in the fugacity (right panel). For portal couplings of $\lambda_{h\phi} \lesssim 1\times10^{-14}$, the number-changing processes do not become efficient, and the dark sector remains dilute throughout the considered temperature range. The solutions ending in filled circles reach the spinodal point where symmetry is restored, while the gray curve at $\lambda_{h\phi}=4.6\times10^{-14}$ marks the boundary where the system does not reach the spinodal point and remains in the broken phase.

\begin{figure}[t!]
    \centering
    \includegraphics[width=0.75\textwidth]
    {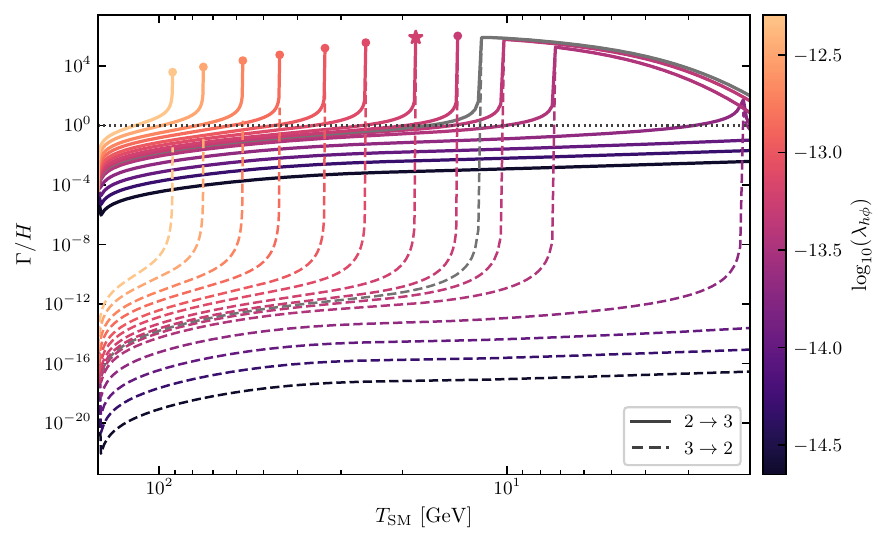}
    \caption{Rates of the number-changing reactions relative to the Hubble expansion for the same portal-coupling scan as in figure~\ref{fig:thermodynamic_evolution}. Solid and dashed curves denote the effective $2\to3$ and $3\to2$ rates, respectively, while the horizontal dotted line marks $\Gamma/H=1$. Colours indicate $\lambda_{h\phi}$. }
    \label{fig:number_changing_rates}
\end{figure} 

The corresponding energy-density evolution is shown in figure~\ref{fig:energy_evolution}. As a reference scale, we indicate the energy density at which the broken and symmetric minima are degenerate in chemical equilibrium,
\begin{align}\label{eq:}
    \rho_{\mathrm{deg}}(z=1)
    \equiv
    \rho_{\mathrm{br}}(T_D^{\mathrm{deg}}, z=1)
    =
    \rho_{\mathrm{sym}}(T_D^{\mathrm{deg}}, z=1)\,,
\end{align}
as the dashed horizontal line. This provides the minimum energy density required for an inverse transition in chemical equilibrium. The trajectories, however, show that reaching this energy scale is not sufficient to trigger symmetry restoration. In fact, the energy density of the symmetry-restoring trajectories typically rises well above $\rho_{\mathrm{deg}}(z=1)$ before the spinodal point is reached. For most values of $\lambda_{h\phi}$, symmetry restoration occurs only later, after the number-changing reactions have redistributed the dark-sector energy into a more highly occupied population of lower-energy modes, which enhances the thermal corrections to the effective potential. While $T_D$ and $z$ exhibit sharp changes during chemical equilibration, as seen in figure~\ref{fig:thermodynamic_evolution}, the energy density $\rho_D$ evolves smoothly. This reflects the fact that the internal number-changing reactions redistribute the available energy without changing the energy density.

To further illustrate the role of the self-number-changing reactions in the phase transition, we display their rates in figure~\ref{fig:number_changing_rates}, normalised to the Hubble expansion rate. In the dilute regime, the reactions are suppressed. When $\Gamma_{2\to3}/H$ becomes larger than unity, the fugacity can change on a time scale shorter than one Hubble time, producing the sharp rise of the rates. We also show the rates $\Gamma_{3\to2}/H$.

\begin{figure}[t!]
    \centering
    \includegraphics[width=0.86\textwidth]
    {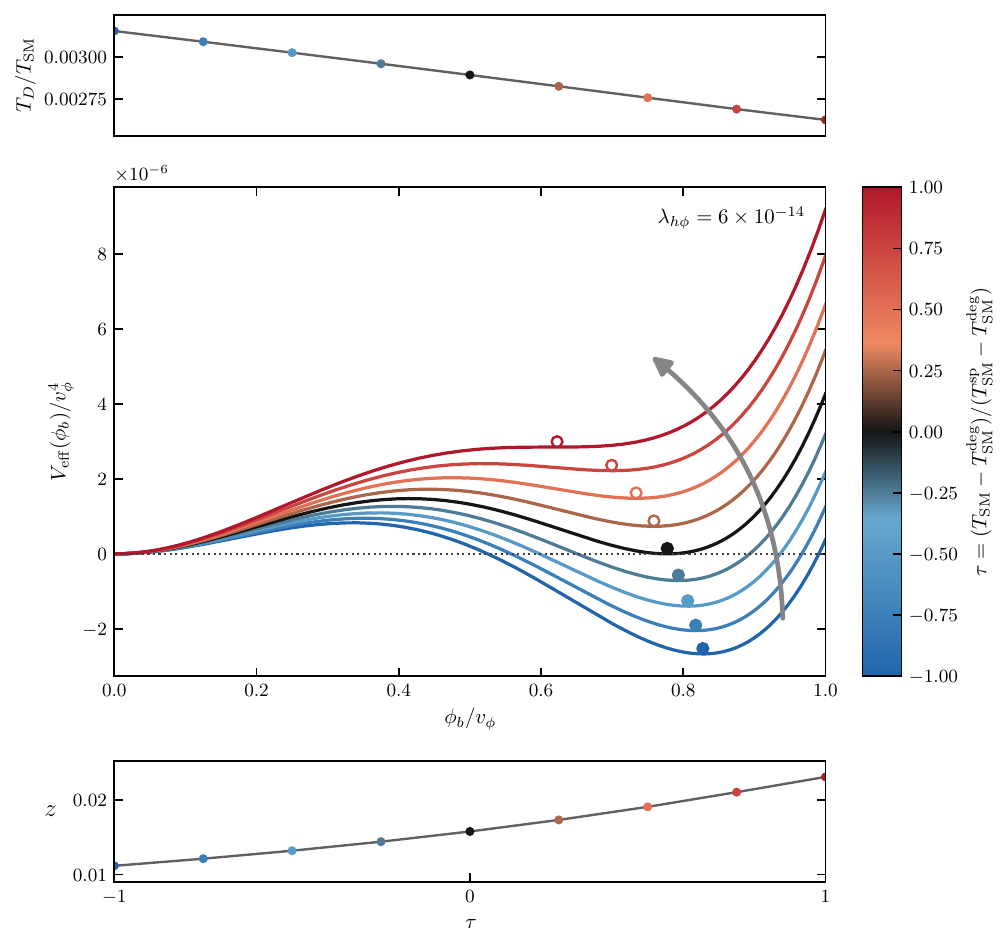}
    \caption{Effective potential for $\lambda_{h\phi}=6\times10^{-14}$, normalised to $v_\phi^4$. The colours indicate the rescaled Standard Model temperature $\tau$ defined in the text, with the black curve corresponding to degeneracy, $\tau=0$. The upper and lower panels show, respectively, $T_D/T_{\rm SM}$ and $z$ over the same interval. Their coloured points correspond one-to-one with the potential curves. Filled circles indicate a globally stable or degenerate broken minimum, whereas hollow circles indicate the metastable broken minimum after degeneracy.}
    \label{fig:transition_potential}
\end{figure}

The change in the phase structure is shown in figure~\ref{fig:transition_potential} for $\lambda_{h\phi}=6\times10^{-14}$. We parametrise the position along the transition interval by $\tau=(T_{\rm SM}-T_{\rm SM}^{\rm deg})/(T_{\rm SM}^{\rm sp}-T_{\rm SM}^{\rm deg})$, where $T_{\rm SM}^{\rm deg}$ is the SM temperature at which the symmetric and broken minima are degenerate (i.e.\ the critical temperature), and $T_{\rm SM}^{\rm sp}$ is the spinodal temperature when the broken phase becomes unstable. For $\tau<0$, the broken minimum is the global minimum. At $\tau=0$, it is degenerate with the symmetric minimum at $\phi_b=0$, while a barrier separates the two phases. For $\tau>0$, the symmetric configuration becomes energetically favoured, and it marks the beginning of the transition. The displayed interval lies around $T_{\rm SM}\simeq18.3\,\mathrm{GeV}$ and spans only about $4.7\,\mathrm{keV}$ in Standard Model temperature, reflecting the rapid redistribution driven by equilibration.

\subsection{Results}

\begin{figure}
    \centering
    \includegraphics[width=1\linewidth]{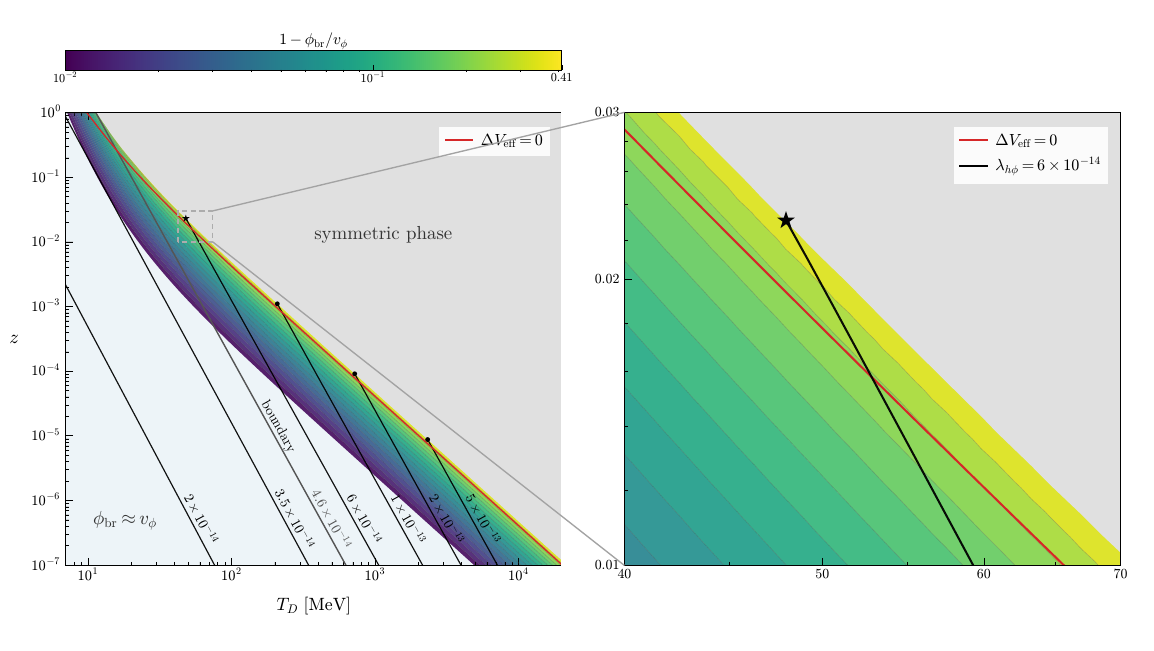}
    \caption{Trajectories in the $(T_D, z)$ phase diagram corresponding to the solutions shown in figure~\ref{fig:thermodynamic_evolution}. As in figure~\ref{fig:vev_scan_perturbativity}, the pale blue-green area highlights the broken phase, while the light-gray area highlights the symmetric phase. The red solid curve denotes the broken-symmetric degeneracy boundary. The evolution proceeds generally from the high-$T_D$, low-$z$ region towards smaller temperature and larger fugacity due to $2\to3$ processes cooling the dark sector. If a trajectory crosses the phase boundary, the dark sector may experience an inverse (symmetry-restoring) phase transition. In such cases we use filled circles to mark the spinodal points when the broken minimum becomes unstable. For the trajectory ending in a star we show the evolution of the effective potential in figure~\ref{fig:transition_potential}.}
\label{fig:phase_diagram_and_trajectories}
\end{figure}

Having obtained the dark sector temperature and fugacity as a function of the scale factor, we can derive the trajectories of the dark sector in the $(T_D,z)$ plane and compare the results with the phase diagram derived in section~\ref{sec:phase_structure}. The result is shown in figure~\ref{fig:phase_diagram_and_trajectories}, with different lines corresponding to different Higgs portal couplings. For the chosen range of fugacities the initial freeze-in production is not visible, i.e.\ the trajectories correspond to the period of rapid evolution towards chemical equilibrium with decreasing dark sector temperature and increasing fugacity. 

In the dilute relativistic limit, the energy density scales as $\rho_D\propto zT_D^4$, such that rapid chemical equilibration with nearly constant $\rho_D$ implies $z \propto T_D^{-4}$. The leading symmetry-restoring term in the effective potential, on the other hand, scales as $zT_D^2$, such that the phase boundary corresponds to $z \propto T_D^{-2}$. Hence, number-changing reactions can substantially enhance the restoring thermal corrections by redistributing the available dark-sector energy into a more highly occupied state, even while cosmological expansion reduces $\rho_D$. This makes it possible for the dark sector to cross the phase boundary and experience an inverse phase transition.

We emphasise that for dark sectors beyond chemical equilibrium, symmetry restoration cannot in general be characterised by a single critical value of the dark-sector energy density. Indeed, the different trajectories in figure~\ref{fig:phase_diagram_and_trajectories} all correspond to different energy densities (with lines further to the right corresponding to larger portal couplings and hence larger energy densities). In each case symmetry restoration happens when the fugacity (and hence the occupation number) becomes large enough that the trajectory crosses the broken--symmetric coexistence boundary.

\subsection{Cosmological constraints}
\label{sec:constraints}

In the discussion up to now, we have not considered the evolution of the dark sector after the inverse phase transition. It is clear, however, that the dark sector cannot remain in the symmetric phase. Once chemical equilibrium has been reached, the expansion of the universe will cause the dark sector to cool down, leading to an ordinary (symmetry-breaking) phase transition. Following this phase transition, all dark sector particles are massive and may contribute to the dark matter of the universe. In this section, we discuss the resulting cosmological constraints.

First of all, we estimate the late-time abundance of dark sector particles for the benchmark point
(table~\ref{tab:benchmark}), for which the vacuum masses are $m_\phi=1.83\,{\rm MeV}$ and
$m_{A'}=30\,{\rm MeV}$. Since $m_{A'}\gg m_\phi$, we expect that the heavier dark photons are
eventually depleted through processes such as $A'A'\to\phi\phi$, leaving the dark Higgs bosons as the
surviving relic. Since this $2\to2$ conversion conserves particle number, the total comoving
dark-sector particle number provides a first estimate of the dark Higgs boson abundance. For example, for the
trajectory with $\lambda_{h\phi}=6\times10^{-14}$, the spinodal endpoint gives
$f_\phi^{\rm end}\equiv\Omega_\phi/\Omega_{\rm DM}\simeq1.8\times10^{-5}$. This is not yet a final relic abundance since
number-changing reactions may be efficient both in the symmetric and in the broken phase. If the inverse phase transition occurs while the dark sector is still out of chemical equilibrium ($z<1$), the expectation is that number-changing processes increase the dark sector particle number. To study this effect in detail, one would need to consider also $2\leftrightarrow 4$ reactions in the symmetric phase, which are beyond the scope of the present work. Instead, we can estimate how chemical equilibration changes the total particle number by assuming energy conservation. The resulting dark matter fraction differs only by a factor of a few from the estimate above and gives
$f_\phi^{\rm late}\simeq(4\text{--}5)\times10^{-5}$.

While it should be clear that the total amount of energy density in the dark sector is small enough to be consistent with the measured dark matter abundance, even a small fraction of dark matter may be constrained by observations. This is because the dark Higgs bosons are not perfectly stable. The assumed Higgs portal coupling leads to a mixing between dark Higgs bosons and SM Higgs bosons with mixing angle
$\theta\simeq\lambda_{h\phi}v_\phi v_h/m_h^2$, which induces dark Higgs boson decays such as $\phi \to e^+ e^-$. While the resulting lifetime is larger than the age of the universe, the decay products may nevertheless be observable.

As an approximate indirect-detection check, we use the dark matter $e^+e^-$ decay bound from ref.~\cite[figure 2]{Cervantes_2024} at $m_\phi\simeq1.83\,{\rm MeV}$. For the trajectory with
$\lambda_{h\phi}=6\times10^{-14}$, we obtain a total width of
$\Gamma_\phi\simeq1.6\times10^{-48}\,{\rm GeV}$, dominated by the $e^+e^-$ channel. The bound in ref.~\cite{Cervantes_2024}
corresponds to $\Gamma_\phi^{\rm full}\simeq2.8\times10^{-49}\,{\rm GeV}$ but assumes that the decaying particle constitutes all of dark matter. Since the decay signal scales as
$\Phi_{\rm decay}\propto f_\phi\Gamma_\phi$, the appropriate constraint for a subdominant component is
$f_\phi\Gamma_\phi\lesssim\Gamma_\phi^{\rm full}$, which is easily satisfied for $\lambda_{h\phi}=6\times10^{-14}$ given our estimate of $f_\phi$ from above.

To determine the largest Higgs portal coupling consistent with the indirect detection constraints, we note that increasing $\lambda_{h\phi}$ increases both the decay width of the dark Higgs boson and the freeze-in production of dark sector states, which also affects the timing of the inverse phase transition. For larger
$\lambda_{h\phi}$, the trajectories reach the spinodal earlier during freeze-in, so that an increasing fraction of the final abundance is produced after symmetry restoration. For a rough estimate we can assume that the final abundance of dark Higgs bosons is proportional to the amount of energy transferred from the SM to the dark sector, which scales with $\lambda_{h\phi}^2$. Taking into account both the change in the lifetime and the change in the abundance, the indirect detection constraint is therefore expected to scale approximately as $\lambda_{h\phi}^4$. We therefore conclude that indirect detection constraints become relevant towards the
upper end of the explored parameter range, $\lambda_{h\phi}\sim{\rm few}\times10^{-13}$. Determining a precise exclusion boundary in this regime would require following the phase transition and subsequent dark-sector evolution in the symmetric phase, which is beyond the scope of this work.

\section{Discussion and conclusions}
\label{sec:conclusions}

We have studied the coupled temperature, chemical potential and phase evolution of a dark $U(1)'$ sector initially in the broken phase and populated through freeze-in. The departure from chemical equilibrium was described in terms of the fugacity $z$, evolved together with the dark-sector temperature $T_D$. By calculating the effective potential in terms of both quantities, we showed that freeze-in alone is insufficient to restore the symmetry, even if the energy provided by freeze-in is in principle enough.  This is because the contribution of a thermal bath with few highly energetic particles (high $T_D$, small $z$) to the effective potential is suppressed compared to a thermal bath in which the same energy density is distributed across more low-energy particles.
Symmetry restoration thus only occurs once the system approaches chemical equilibrium through number-changing processes. We have studied this evolution using coupled Boltzmann equations together with the evolving background field in order to correctly account for the thermal masses.

Our results show that the phase history cannot, in general, be inferred from the dark-sector temperature alone. As shown in figure~\ref{fig:phase_diagram_and_trajectories}, symmetry restoration is instead controlled by the combined evolution of $T_D$ and $z$ and occurs when the trajectory crosses the degeneracy boundary and approaches the spinodal configuration. Treating the dark sector as chemically equilibrated from the outset would miss the redistribution stage and could therefore shift the transition boundary or predict a transition
that does not actually complete.

For the model and benchmark point that we have considered, chemical equilibration occurs in a narrow cosmological time window as soon as the number-changing processes become efficient, leading to a very fast phase transition. The dynamics of such a rapid transition would effectively be described by a second-order transition, where the system reaches the spinodal point and the background field continuously rolls into the global minimum. If the evolution proceeded more slowly, the barrier in the effective potential would create the conditions for a first-order phase transition from the metastable to the true minimum. The expanding bubbles of the true vacuum could then transfer their energy to the surrounding plasma and source gravitational waves, as studied extensively for direct transitions~\cite{Hindmarsh_2015,Hindmarsh:2016lnk,Barni_2024,Caprini:2024hue} and recently also for inverse transitions~\cite{Barni_2024,Barni:2025gnm}. In standard non-runaway plasma transitions, the resulting sound waves are expected to provide the leading contribution to the gravitational-wave spectrum~\cite{Hindmarsh_2015,Hindmarsh:2016lnk,Barni_2024,Caprini:2024hue}.

If chemical equilibration happens rapidly, we expect an
approximately exponential bubble nucleation rate $\Gamma\propto e^{\beta t}$ with $\beta\gg H$~\cite{Turner:1992tz,Hogan:1983ixn}, which schematically gives $N_{\rm bub}\sim\Gamma/(\beta H^3)$. One bubble per Hubble volume therefore corresponds to $\Gamma/H^4\sim\beta/H$. If nucleation and percolation both occur during this rapid stage, the resulting $H_*/\beta$ is small~\cite{Caprini_2016}, leading to a suppressed gravitational-wave signal. Moreover, only a small fraction of the total energy density is stored in the dark sector, additionally limiting the effective transition strength $\alpha_{\rm tot}$ and hence the achievable energy density in gravitational waves.

To increase the gravitational-wave signal, it appears necessary to consider scenarios where chemical equilibration occurs more slowly over a longer period of time, either because number-changing processes are less efficient or because freeze-in production proceeds in a different way than assumed here. For instance, kinetic mixing predicts a more complicated energy transfer history, including resonant conversion, which could lead to a more energetic dark plasma. Slow decays of a long-lived scalar condensate could instead sustain particle injection over many Hubble times and allow the system to approach the phase boundary more gradually via slower chemical equilibration. Either possibility could increase $\alpha_{\rm tot}$, reduce $\beta/H_*$, or provide more time for nucleation and bubble growth. Studying the nucleation history and gravitational-wave signal in such scenarios is a promising direction for future work. Studying first-order phase transitions induced by freeze-in offers the exciting possibility of testing dark sectors with feeble couplings to matter.

\acknowledgments

We thank Andrzej Hryczuk, Marek Lewicki and Thomas Schwetz for useful discussions. EC is supported by the
National Science Centre (Poland) under the research Grant No. 2021/42/E/ST2/00009. This research was supported by the Deutsche Forschungsgemeinschaft (DFG) through Grant No.\ 396021762 -- TRR 257.



\appendix

\section{Convergence of the fugacity expansion}
\label{app:kf_convergence}

\begin{figure}[t!]
    \centering
    \includegraphics[width=0.65\textwidth]
    {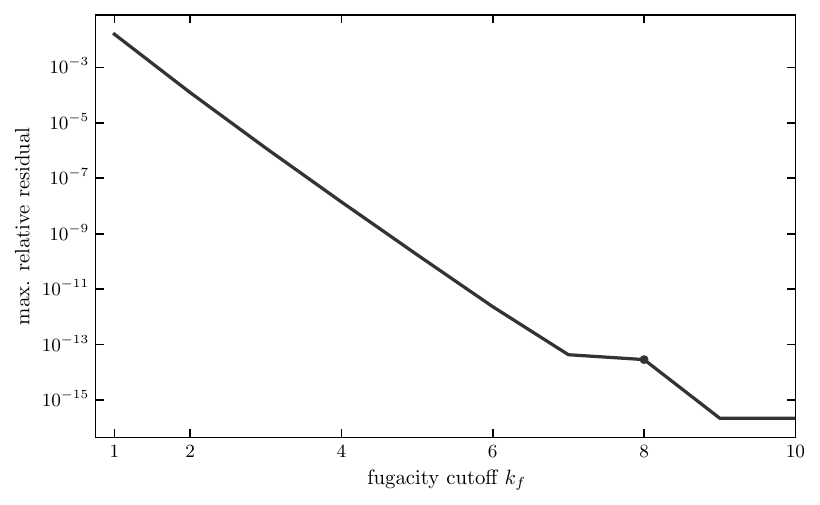}
    \caption{Maximum relative residual of the screened effective potential at the near-spinodal endpoint of the $\lambda_{h\phi}=6\times10^{-14}$ benchmark. The marker indicates the cutoff $k_f=8$ used in the numerical evolution. Only $k_f\leq10$ is shown, since the residual has already reached the numerical precision floor at larger cutoffs.}
    \label{fig:kf_convergence}
\end{figure}

Figure~\ref{fig:kf_convergence} shows the convergence of the screened effective potential for the benchmark $\lambda_{h\phi}=6\times10^{-14}$ at the numerical near-spinodal endpoint, where the Bose--Einstein corrections are largest. Writing $V_{k_f}\equiv V_{\rm eff}^{(k_f)}$, we define the maximum relative residual with respect to the $k_f=20$ result as
\begin{equation}
\mathcal R(k_f)=\frac{\max_\phi\left|V_{k_f}(\phi)-V_{20}(\phi)\right|}{\max\left\{\max_\phi V_{20}-\min_\phi V_{20},\,\max_\phi|V_{20}|\right\}}\,.
\end{equation}
The $k_f=20$ result is retained as the reference value, while only $k_f\leq10$ is displayed since increasing the cutoff further does not lead to an improvement in double precision. At the cutoff $k_{\max}=8$ used in our calculation, $\mathcal R(8)\simeq3\times10^{-14}$.

\section{Screening}
\label{app:finite_z_screening}

In this appendix we provide the derivation of the screened medium prescription
used in section~\ref{sec:screened_V}. The appropriate starting point is the
tadpole, or equivalently the derivative of the medium contribution with
respect to the background field. This makes explicit the relation between
partial dressing and the screened-mass prescription employed in the main
text.

For a bosonic species or polarization sector $i$ with $g_i$ degrees of
freedom, we defined
\begin{equation}
\mathcal V_i(m_i^2;T_D,z)
=
-P_i(m_i^2;T_D,z),
\end{equation}
where $P_i$ is given in eq.~\eqref{eq:species_pressure}. The tadpole
$\mathcal I_i$ in eq.~\eqref{eq:tadpole_integral} is defined per bosonic
degree of freedom. Hence
\begin{equation}
\frac{\partial\mathcal V_i}{\partial m_i^2}
=
\frac{g_i}{2}\,
\mathcal I_i(m_i;T_D,z)=\frac{g_i}{2}\int_p \frac{f_i}{E_i}.
\label{eq:app_medium_tadpole_identity}
\end{equation}
Using $m_i^2=m_i^2(\phi_b)$, the corresponding one-loop medium force is
therefore
\begin{equation}
\frac{{\rm d}\mathcal V_i}{{\rm d}\phi_b}
=
\frac{g_i}{2}
\frac{{\rm d}m_i^2}{{\rm d}\phi_b}
\mathcal I_i(m_i;T_D,z).
\label{eq:app_undressed_force}
\end{equation}

The same tadpole integral determines the leading static self-energies. In the real-time formalism, the medium part of the bosonic propagator is $iD^{11}_{\rm med}(p)=2\pi\delta(p^2-m^2)f(p)$. A one-loop tadpole generated by a quartic interaction therefore has the generic form $\Pi_i=c_i\int {\rm d}^4p/(2\pi)^4\, 2\pi\delta(p^2-m_i^2)f_i(p)$, where $c_i$ contains the coupling and combinatorial factors. Performing the $p^0$ integral gives $\Pi_i=c_i\int_p f_i(p)/E_i=c_i\mathcal I_i$. For the interactions of the $U(1)'$ model, $\mathcal L_{\rm int}\supset -\lambda\phi^4/24-\lambda\phi^2\varphi^2/12-\lambda\varphi^4/24 +g^2A'_\mu A'^\mu(\phi^2+\varphi^2)/2$, the scalar tadpole symmetry factors give the coefficients $\lambda/2$ and $\lambda/6$ in eq.~\eqref{eq:thermal_self_energies}, while the gauge seagull gives $3g^2\mathcal I_{A'}$. For the longitudinal gauge mode, the scalar bubble and seagull contributions to the static polarization $\Pi_{00}(0,\mathbf 0)$ combine to give $2g^2\mathcal I_j$ for each real scalar component.

Partial dressing amounts to evaluating the tadpole with a screened
propagator while leaving the explicit background derivative of the
tree-level mass unchanged~\cite{Boyd:1993tz}. Introducing
\begin{equation}
M_i^2(\phi_b;T_D,z)
=
m_i^2(\phi_b)+\Pi_i(\phi_b;T_D,z),
\end{equation}
the partially dressed force is
\begin{equation}
\frac{{\rm d}\mathcal V_i^{\rm PD}}{{\rm d}\phi_b}
=
\frac{g_i}{2}
\frac{{\rm d}m_i^2}{{\rm d}\phi_b}
\mathcal I_i(M_i;T_D,z).
\label{eq:app_dressed_force}
\end{equation}
The screened potential is obtained from this force by integration over the
background field,
\begin{equation}
\mathcal V_i^{\rm PD}(\phi_b;T_D,z)
-
\mathcal V_i^{\rm PD}(\phi_{b,0};T_D,z)
=
\frac{g_i}{2}
\int_{\phi_{b,0}}^{\phi_b}
{\rm d}\varphi\,
\frac{{\rm d}m_i^2(\varphi)}{{\rm d}\varphi}
\mathcal I_i\!\left(M_i(\varphi);T_D,z\right).
\label{eq:app_partial_dressing_integral}
\end{equation}
Equation~\eqref{eq:app_partial_dressing_integral} is the general
background-field form of the partial-dressing prescription.

The previous result can be simplified because we evaluate the screening self-energies in the relativistic approximation, $\mathcal I_i(T_D,z)
=
\frac{T_D^2}{2\pi^2}
\operatorname{Li}_2(z),$
so that the leading self-energies depend on $T_D$ and $z$, but not on
$\phi_b$. Then
\begin{equation}
\frac{{\rm d}M_i^2}{{\rm d}\phi_b}
=
\frac{{\rm d}m_i^2}{{\rm d}\phi_b}.
\end{equation}
Together with eq.~\eqref{eq:app_medium_tadpole_identity}, this implies
\begin{equation}
\frac{{\rm d}}{{\rm d}\phi_b}
\mathcal V_i(M_i^2;T_D,z)
=
\frac{g_i}{2}
\frac{{\rm d}m_i^2}{{\rm d}\phi_b}
\mathcal I_i(M_i;T_D,z).
\end{equation}
The integrand in eq.~\eqref{eq:app_partial_dressing_integral} is therefore
an exact derivative, and
\begin{equation}
\mathcal V_i^{\rm PD}(\phi_b;T_D,z)
=
\mathcal V_i(M_i^2;T_D,z)
+
Q_i(T_D,z),
\label{eq:app_pd_mass_replacement}
\end{equation}
where $Q_i$ is independent of the background field. Thus, within the relativistic self-energy approximation used in this work, no additional numerical integration over $\phi_b$ is required. Partial dressing is equivalent, up to a background-independent contribution, to the screened mass replacement used in eq.~\eqref{eq:screened_medium_potential}. For the scalar modes this follows directly from the partial-dressing construction. For the longitudinal gauge mode, the corresponding use of the distribution-dependent static screening mass should be understood as the leading quasiparticle screening prescription motivated by the real-time resummed propagator~\cite{Carrington:1997sq}. The two transverse gauge polarizations remain unscreened at this order.

Let us briefly discuss perturbativity for the case of a dilute dark sector. An estimate of perturbative control in a  bosonic plasma in chemical equilibrium is provided by $\epsilon_{\rm eq}\sim cT_D/(\pi M_i)$, where $c$ denotes the relevant coupling and $M_i$ the screened mass of the corresponding bosonic mode~\cite{Laine_2016}. This estimate reflects the classical $T_D/E_i$ enhancement of soft equilibrium modes and the necessity of daisy resummation when $\epsilon_{\rm eq}\gtrsim 1$. Indeed, for
$E\ll T_D$, the equilibrium distribution function can be approximated as
\begin{equation}
\left.\frac{f(E)}{E}\right|_{z=1}
\simeq
\frac{T_D}{E^2} \; ,
\end{equation}
which produces the power-like infrared enhancement underlying the usual three-dimensional soft-mode expansion. For $z< 1$, however, the soft occupation is suppressed by $\epsilon\propto z$, alleviating the infrared breakdown. The distribution function can then be approximated as
\begin{equation}
\frac{f(E)}{E}
\xrightarrow[E\to0]{}
\frac{z}{1-z}\frac{1}{E} \; ,
\end{equation}
so the parameter controlling the infrared divergence is proportional to $z$. The occupation of the softest modes remains suppressed by $z$, rather than growing as $T_D/E$, and the soft enhancement is absent when $z<1$.

As a consistency check, we consider the chemical-equilibrium case. For $z=1$ and $m_i/T_D\ll1$, the medium contribution of a bosonic sector with
$g_i$ degrees of freedom is
\begin{equation}
\mathcal V_i(m_i^2;T_D,1)
=
-g_i\frac{\pi^2T_D^4}{90}
+
g_i\frac{m_i^2T_D^2}{24}
-
g_i\frac{T_D}{12\pi}
\left(m_i^2\right)^{3/2}
+\mathcal O\left(
m_i^4\log\frac{m_i^2}{T_D^2}
\right).
\label{eq:app_equilibrium_expansion}
\end{equation}
Replacing $m_i^2\rightarrow M_i^2=m_i^2+\Pi_i$ gives the residual
\begin{equation}
\begin{aligned}
\mathcal V_i(M_i^2;T_D,1)
-\mathcal V_i(m_i^2;T_D,1)
=
g_i\frac{T_D^2\Pi_i}{24}
-
g_i\frac{T_D}{12\pi}
\left[
\left(m_i^2+\Pi_i\right)^{3/2}
-\left(m_i^2\right)^{3/2}
\right]
+\cdots .
\end{aligned}
\label{eq:app_equilibrium_ring_limit}
\end{equation}
The leading field-dependent contribution is precisely the usual Arnold--Espinosa ring term~\cite{Arnold:1992rz}, recovering the conventional equilibrium resummation in the high-temperature limit. In contrast, the relativistic expansion is qualitatively different at $z<1$. Expanding the expression in eq.~\eqref{eq:species_pressure} gives
\begin{equation}
\begin{aligned}
\mathcal V_i(m_i^2;T_D,z)
={}&
-g_i\frac{T_D^4}{\pi^2}\operatorname{Li}_4(z)
+
g_i\frac{m_i^2T_D^2}{4\pi^2}\operatorname{Li}_2(z)
\\
&+
g_i\frac{m_i^4}{32\pi^2}
\left[
\frac{z}{1-z}
\log\left(\frac{m_i^2}{T_D^2}\right)
+
c(z)
\right]
+\cdots ,
\qquad 0<z<1 ,
\end{aligned}
\label{eq:app_fixed_z_expansion}
\end{equation}
where $c(z)$ is independent of $m_i$ and collects the analytic contribution at order $m_i^4$. In particular, at any $z<1$ there is no $m_i^3T_D$ term as in the $z=1$ case. The first non-analytic mass dependence is instead proportional to $m_i^4\log m_i^2$.

\section{Number-changing collision terms}
\label{app:collision_operators}

For a channel $r$: $a+b\to c+d+e$, we define $\gamma_r^{\rm eq}$ as the density of forward reactions when all species have equilibrium Maxwell-Boltzmann distributions\footnote{The Maxwell-Boltzmann approximation is adopted only for the collision integrals; in the underoccupied regime relevant for the transition, quantum-statistical corrections to the reaction densities are suppressed by additional powers of the phase-space occupation, while full Bose--Einstein statistics is retained in the thermodynamic quantities entering the effective potential.
} with the common dark-sector temperature $T_D$.  In terms of the unpolarized tree-level cross section, this reaction density is
\begin{equation}
 \gamma_r^{\rm eq}(T_D)
 =
 \mathcal S_r\frac{T_D}{32\pi^4}
 \int_{s_{\min,r}}^\infty \!{\rm d}s\,
 \frac{\lambda_K(s,m_a^2,m_b^2)}{\sqrt{s}}
 K_1\!\left(\frac{\sqrt{s}}{T_D}\right)
 \sigma_r(s),
 \label{eq:gamma_eq_thermal_average}
\end{equation}
where $\lambda_K(x,y,z)=x^2+y^2+z^2-2xy-2xz-2yz$, $s_{\min,r}=\max\{(m_a+m_b)^2,(m_c+m_d+m_e)^2\}$, and $K_1$ is a modified Bessel function.  The factor $\mathcal S_r$ contains the initial-state degeneracies and the symmetry factor for identical incoming particles; with cross sections averaged over the incoming internal states, $\mathcal S_r=g_ag_b/(1+\delta_{ab})$.  Symmetry factors for identical final states are already contained in the cross sections generated with \texttt{CalcHEP}~\cite{Belyaev:2012qa}.  Equation~\eqref{eq:gamma_eq_thermal_average} has mass dimension four, as required for the zeroth moment of the number-density collision operator, and corresponds to the standard Lorentz-invariant representation of a thermal reaction density, generalized here to the corresponding $2\to3$ case \cite{Gondolo:1990dk}.

\begin{figure}[t]
    \centering
    \includegraphics[width=0.72\textwidth]
    {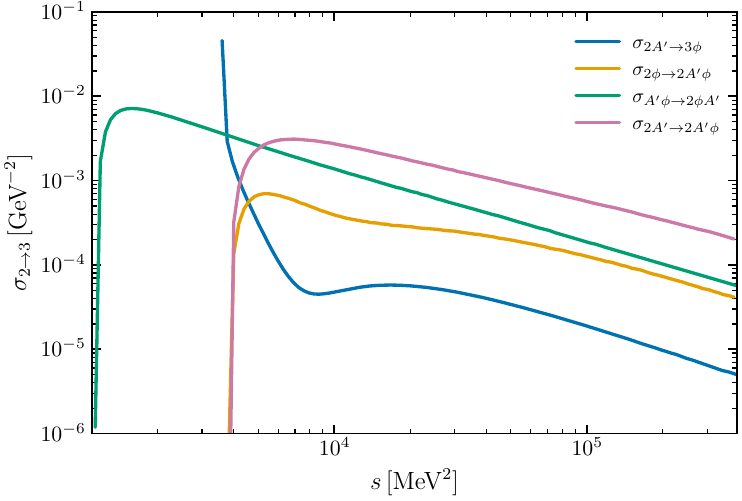}
    \caption{Tree-level broken-phase cross sections for the four gauge-assisted $2\to3$ channels retained in the numerical collision sum. }
    \label{fig:two_to_three_cross_sections}
\end{figure}

The collision term is evaluated in the common-fugacity approximation.  Each incoming particle supplies one power of $z$, while detailed balance makes the forward and inverse equilibrium reaction densities equal.  The realized reaction densities are therefore
\begin{equation}
 \gamma_{r,2\to3}(T_D,z)=z^2\gamma_r^{\rm eq}(T_D),
 \qquad
 \gamma_{r,3\to2}(T_D,z)=z^3\gamma_r^{\rm eq}(T_D).
 \label{eq:realized_23_reaction_densities}
\end{equation}
Since a forward event increases the total dark-particle number by one and an inverse event decreases it by one, the contribution of channel $r$ to the
number-density equation is
\begin{equation}
 C_{0,r}^{\rm nc}
 =\gamma_{r,2\to3}-\gamma_{r,3\to2}
 =\left(z^2-z^3\right)\gamma_r^{\rm eq}.
 \label{eq:channel_number_changing_source}
\end{equation}
Summing eq.~\eqref{eq:channel_number_changing_source} over all possible processes gives
eq.~\eqref{eq:total_number_changing_source}.
Figure~\ref{fig:two_to_three_cross_sections} shows the four gauge-assisted channels. These cross sections are thermally averaged according to eq.~\eqref{eq:gamma_eq_thermal_average}; the scalar-only cross section satisfies $\sigma_{2\phi\to3\phi} \lesssim 3\times10^{-11}\,{\rm GeV}^{-2}$ for all centre-of-mass energies, and is therefore negligible and not displayed. 

Finally, defining $\gamma_{\rm tot}^{\rm eq}=\sum_r\gamma_r^{\rm eq}$ and denoting the total dark-sector number density by $n_D$, the two rates displayed in figure~\ref{fig:number_changing_rates} are
\begin{equation}
 \Gamma_{2\to3}=\frac{z^2\gamma_{\rm tot}^{\rm eq}}{n_D},
 \qquad
 \Gamma_{3\to2}=\frac{z^3\gamma_{\rm tot}^{\rm eq}}{n_D}.
 \label{eq:realized_rates}
\end{equation}
Thus, the curves in that figure show the actually realized forward and inverse rates, not an equilibrium estimate.

\section{Further details on the temperature and fugacity evolution}
\label{app:scale_factor_evolution}

In this appendix, we provide expressions for the partial derivatives of the dark sector number and energy density, $n_D$ and $\rho_D$, with respect to the dark sector temperature and fugacity, $T_D$ and $z$, in terms of the effective potential $V_\text{eff}$. The homogeneous background follows the minimum $\phi_{\min}(T_D,z)$ defined in eq.~\eqref{eq:instantaneous_minimum}. Differentiating the minimum condition gives
\begin{equation}
\frac{\partial\phi_{\min}}{\partial X}
=
-\frac{V_{\phi X}}{V_{\phi\phi}},
\qquad
X\in\{T_D,z\},
\label{eq:minimum_response}
\end{equation}
where in this appendix $V_{\phi X}$, $V_{\phi\phi}$, and analogous quantities denote partial derivatives of $V_{\rm eff}(\phi_b;T_D,z)$ evaluated at $\phi_b=\phi_{\min}(T_D,z)$.
Let us now consider the branch-restricted potential $\tilde V(T_D,z)$ defined in eq.~\eqref{eq:reduced_effective_potential}. Its first derivatives satisfy
\begin{equation}
\tilde V_X
=
V_X,
\qquad
X\in\{T_D,z\},
\label{eq:reduced_potential_first_derivatives}
\end{equation}
because $V_\phi=0$ at the minimum. Its second derivatives are instead
\begin{equation}
\tilde V_{XY}
=
V_{XY}
-
\frac{V_{\phi X}V_{\phi Y}}{V_{\phi\phi}},
\qquad
X,Y\in\{T_D,z\}.
\label{eq:reduced_potential_second_derivatives}
\end{equation}
The second term therefore accounts for the displacement of the minimum as $T_D$ or $z$ is varied. Using the thermodynamic definitions in eq.~\eqref{eq:thermodynamics_from_reduced_potential},
\begin{equation}
P_D=-\tilde V,
\qquad
\rho_D=\tilde V-T_D\tilde V_T,
\end{equation}
the derivatives of the energy density along the minimum are
\begin{align}
\frac{\partial \rho_D}{\partial T_D}
&\equiv
\left(\frac{\partial\rho_D}{\partial T_D}\right)_z
=
-T_D\tilde V_{TT}
=
T_D
\left(
\frac{V_{\phi T}^2}{V_{\phi\phi}}
-
V_{TT}
\right),
\label{eq:energy_temperature_derivative}
\\
\frac{\partial \rho_D}{\partial z}
&\equiv
\left(\frac{\partial\rho_D}{\partial z}\right)_{T_D}
=
\tilde V_z
-
T_D\tilde V_{Tz}
 =
V_z
-
T_DV_{Tz}
+
T_D
\frac{V_{\phi T}V_{\phi z}}{V_{\phi\phi}} \; .
\label{eq:energy_fugacity_derivative}
\end{align}
Here we use a subscript $T$ to denote differentiation with respect to $T_D$.

Writing the physical number density as in eq.~\eqref{eq:physical_quasiparticle_number}, its derivatives along the minimum are
\begin{align}
\frac{\partial n_D}{\partial T_D}
&\equiv
\left(\frac{\partial n_D}{\partial T_D}\right)_z
=
\sum_i
\left[
\left(\frac{\partial n_i}{\partial T_D}\right)_{z,M_i^2}
+
\left(\frac{\partial n_i}{\partial M_i^2}\right)_{T_D,z}
\frac{{\rm d}M_i^2}{{\rm d}T_D}
\right],
\label{eq:number_temperature_derivative}
\\
\frac{\partial n_D}{\partial z}
&\equiv
\left(\frac{\partial n_D}{\partial z}\right)_{T_D}
=
\sum_i
\left[
\left(\frac{\partial n_i}{\partial z}\right)_{T_D,M_i^2}
+
\left(\frac{\partial n_i}{\partial M_i^2}\right)_{T_D,z}
\frac{{\rm d}M_i^2}{{\rm d}z}
\right].
\label{eq:number_fugacity_derivative}
\end{align}
The sum runs over $i\in\{\phi,A'_T,A'_L\}$, and each $n_i$ already contains the corresponding degeneracy factor $g_i$ as in the main text. No additional factor of $g_i$ is therefore required in eqs.~\eqref{eq:number_temperature_derivative} and \eqref{eq:number_fugacity_derivative}. The Goldstone mode remains part of the Landau-gauge effective potential but is not included in this physical number density. Here ${\rm d}M_i^2/{\rm d}X$ includes the response of $\phi_{\min}(T_D,z)$ from eq.~\eqref{eq:minimum_response}. These terms modify the Jacobian that maps the two physical moments to $(T_D,z)$; they do not add a source to the number equation.

For the radial scalar, the mass entering the number-density sum is the screened propagator mass rather than the curvature of the effective potential,
\begin{align}
M_\phi^2
&=
m_{\phi,{\rm tree}}^2(\phi_{\min})
+\Pi_\phi(T_D,z),
&
\frac{{\rm d}M_\phi^2}{{\rm d}X}
&=
\lambda\phi_{\min}
\frac{\partial\phi_{\min}}{\partial X}
+
\frac{\partial\Pi_\phi}{\partial X},
\nonumber
\\
M_{A'_T}^2
&=
g^2\phi_{\min}^2,
&
\frac{{\rm d}M_{A'_T}^2}{{\rm d}X}
&=
2g^2\phi_{\min}
\frac{\partial\phi_{\min}}{\partial X},
\nonumber
\\
M_{A'_L}^2
&=
M_{A'_T}^2
+
\frac{2g^2T_D^2}{\pi^2}\mathcal L_2(z),
&
\mathcal L_2(z)
&\equiv
\sum_{k=1}^{k_f}\frac{z^k}{k^2},
\label{eq:quasiparticle_mass_responses_app}
\end{align}
with $X\in\{T_D,z\}$. We use $M_{A'_T}^2\equiv m_{A'_T}^2$ for notational uniformity because the transverse mode is unscreened at this order. In the relativistic self-energy approximation of eq.~\eqref{eq:thermal_self_energies},
\begin{align}
\Pi_\phi(T_D,z)
&=
\left(\frac{2\lambda}{3}+3g^2\right)
\frac{T_D^2}{2\pi^2}\mathcal L_2(z),
\nonumber
\\
\frac{\partial\Pi_\phi}{\partial T_D}
&=
\left(\frac{2\lambda}{3}+3g^2\right)
\frac{T_D}{\pi^2}\mathcal L_2(z),
\nonumber
\\
\frac{\partial\Pi_\phi}{\partial z}
&=
\left(\frac{2\lambda}{3}+3g^2\right)
\frac{T_D^2}{2\pi^2}
\sum_{k=1}^{k_f}\frac{z^{k-1}}{k}.
\label{eq:scalar_self_energy_responses_app}
\end{align}
The longitudinal mass derivative includes both the transverse-mass response and the explicit derivative of its screening term. In particular,
\begin{align}
\frac{{\rm d}M_{A'_L}^2}{{\rm d}T_D}
&=
\frac{{\rm d}M_{A'_T}^2}{{\rm d}T_D}
+
\frac{4g^2T_D}{\pi^2}\mathcal L_2(z),
\nonumber
\\
\frac{{\rm d}M_{A'_L}^2}{{\rm d}z}
&=
\frac{{\rm d}M_{A'_T}^2}{{\rm d}z}
+
\frac{2g^2T_D^2}{\pi^2}
\sum_{k=1}^{k_f}\frac{z^{k-1}}{k}.
\label{eq:longitudinal_mass_responses_app}
\end{align}

\bibliographystyle{JHEP}
\bibliography{biblio}
\end{document}